\documentclass{nature3}

\usepackage{journaux}
\usepackage{graphicx}

\usepackage{floatrow}
\usepackage{booktabs,caption}
\usepackage[flushleft]{threeparttable}
\usepackage{array,multirow,makecell}
\usepackage{txfonts}

\title{The unexpectedly large proportion of high-mass star-forming cores in a Galactic mini-starburst}

\author{F. Motte$^{1,2\,\star}$, T. Nony$^{1\,\star}$, F. Louvet$^{3\,\star}$, K. A. Marsh$^4$, S. Bontemps$^5$, A. P. Whitworth$^4$, A. Men'shchikov$^{2}$, Q. Nguyen Luong$^{6,7}$, T. Csengeri$^8$, A. J. Maury$^2$, A. Gusdorf$^9$, E. Chapillon$^{5,10}$, V. K\"onyves$^2$, P. Schilke$^{11}$, A. Duarte-Cabral$^4$, P. Didelon$^2$ \& M. Gaudel$^2$}

\newcommand{\msun}{\rm{M}_\odot}
\newcommand{\lsun}{\rm{L}_\odot}

\begin{document}

\maketitle

\begin{affiliations}
 \item Universit\'e Grenoble Alpes, CNRS, Institut de Plan\'etologie et d'Astrophysique de Grenoble, F-38000 Grenoble, France
 \item AIM Paris-Saclay/D\'epartement d'Astrophysique, CEA, CNRS, Univ. Paris Diderot, CEA-Saclay, F-91191 Gif-sur-Yvette Cedex, France
 \item  Department of Astronomy, Universidad de Chile, Las Condes, Santiago, Chile
 \item School of Physics and Astronomy, Cardiff University, Queens Buildings, The Parade, Cardiff CF24 3AA, UK
 \item OASU/Laboratoire d'Astrophysique de Bordeaux, Univ. Bordeaux, CNRS, F-33615 Pessac, France
 \item  Korea Astronomy and Space Science Institute, 776 Daedeok daero, Yuseoung, Daejeon 34055, Republic of Korea
 \item  NAOJ Chile Observatory, National Astronomical Observatory of Japan, 2-21-1 Osawa, Mitaka, Tokyo 181-8588, Japan
 \item Max-Planck-Institut f\"ur Radioastronomie, Auf dem H\"ugel 69, 53121 Bonn, Germany
 \item LERMA, Observatoire de Paris, PSL Research University, CNRS, Sorbonne Universit\'es, UPMC Univ. Paris 06, \'Ecole normale sup\'erieure, F-75005, Paris, France 
 \item Institut de RadioAstronomie Millim\'etrique (IRAM), 300 rue de la Piscine, F-38406 Saint Martin d'H\`eres, France
 \item I. Physik. Institut, University of Cologne, 50937 K\"oln, Germany
\end{affiliations}


{\bf Understanding the processes that determine the stellar Initial Mass Function (IMF) is a critical unsolved problem, with profound implications for many areas of astrophysics\textsuperscript{\bf \cite{offner14}}. In molecular clouds, stars are formed in cores, gas condensations which are sufficiently dense that gravitational collapse converts a large fraction of their mass into a star or small clutch of stars. In nearby star-formation regions, the core mass function (CMF) is strikingly similar to the IMF, suggesting that the shape of the IMF may simply be inherited from the CMF\textsuperscript{\bf \cite{motte98, TeSa98, enoch08, konyves15}}. Here we present $1.3\,{\rm mm}$ observations, obtained with ALMA, the world's largest interferometer, of the active star-formation region W43-MM1, which may be more representative of the Galactic-disk regions where most stars form\textsuperscript{\bf \cite{nguyen13, louvet14}}. The unprecedented resolution of these observations reveals, for the first time, a statistically robust CMF at high masses, with a slope that is markedly shallower than the IMF.  This seriously challenges our understanding of the origin of the IMF.}

The IMF, giving the relative numbers of stars born with different masses, between $\sim\!0.08\,\msun$ and $\sim\!100\,\msun$, appears to be universal\textsuperscript{\bf \cite{bastian10, kroupa13}}, although there are a few claims to the contrary, especially in young massive clusters\textsuperscript{\bf \cite{harayama08,maia16}}. In nearby star-formation regions, stars with masses between $\sim\!0.08\,\msun$ and $\sim\!5\,\msun$ are observed to form in cores, and the CMF in these  regions has an approximately log-normal shape with a power-law tail at higher masses ($\stackrel{>}{\sim} \msun$), very similar to the IMF, but with the peak shifted to higher masses by a factor $1/\epsilon\!\sim2.5\pm1$\textsuperscript{\bf \cite{motte98, TeSa98,  enoch08, konyves15}}. This suggests that the shape of the IMF is determined by the processes that determine the CMF, in which case there must be an approximately self-similar mapping from core mass to stellar mass, with a very small variance (see Methods). However, nearby star-formation regions are probably not representative of the regions where most stars are formed, {\it and} they do not form massive stars ($\sim\!8\,\msun$ to $\sim\!100\,\msun$)\textsuperscript{\bf \cite{motte18}}. Previous attempts to resolve cores in high-mass star-formation regions have suffered from poor statistics\textsuperscript{\bf \cite{andre10, bontemps10}}, and therefore have not allowed a critical comparison of the CMF and IMF\textsuperscript{\bf \cite{bontemps10, zhang15, cheng18}}.

W43-MM1 is a massive cloud ($\sim\!2\times 10^4\,\msun$) located at the tip of the Galactic bar, a very rich region in terms of cloud concentration and star formation activity\textsuperscript{\bf \cite{motte03, nguyen13, louvet14}}, $\sim\!5.5\,{\rm kpc}$ from the Sun\textsuperscript{\bf \cite{nguyen11b}}. Shock tracing lines like SiO are detected throughout this cloud, testifying to the prevalance of inflowing and colliding gas streams\textsuperscript{\bf \cite{nguyen13, louvet16}}. The large densities in W43-MM1, traced at high angular resolution by millimetre continuum radiation, suggest it will form a massive star cluster in the near future\textsuperscript{\bf \cite{louvet14}}. We have used ALMA to image the complete W43-MM1 cloud ($\sim\!3\,{\rm pc}^2$) at $1.3\,{\rm mm}$ (see Supplementary Fig.~1), with unprecedented resolution ($\sim\! 2,400\,{\rm AU}$), and hence to identify a large sample of individual cores, with a broad and previously unmatched range of masses. 

Since the $1.3\,{\rm mm}$ map of W43-MM1 has a dynamic range of 30 in physical scale, and 4,000 in flux, we use the \emph{getsources} algorithm\textsuperscript{\bf \cite{menshchikov12}} to identify and characterise cores. \emph{getsources} decomposes cloud emission into a multi-resolution data-cube; two cube dimensions give position on the sky, and the third dimension is the physical scale of cloud substructure. Cores are identified at small scales, and their positions, spatial extents (best-fitting ellipses at half-maximum), peak and integrated fluxes (corrected for local background), masses, and mean densities are computed (see Supplementary Table~1). Fig.~1 shows the elliptical outlines of the 131 cores returned by \emph{getsources}, of which 94 are deemed to be very robust (significance, ${\rm Sig}\!\ge\!7$). The cores have diameters ranging from $\sim\!1,300\,{\rm AU}$ to $\sim\!3,500\,{\rm AU}$ once deconvolved with the $0.44''$ (or $\sim\!2,400\,{\rm AU}$) telescope beam; masses ranging from $\sim\!1\,\msun$ to $\sim\!100\,\msun$; and mean H$_{_2}$ densities ranging from $\sim\!10^7\,{\rm cm}^{-3}\;$ to $\;\sim\!10^{10}\,{\rm cm}^{-3}$. Monte-Carlo extraction simulations indicate that the sample is 90\% complete down to $\sim\!1.6\,\msun$ (see Methods).

The blue histogram in Fig.~2a displays the differential CMF of W43-MM1, $dN/d\log(M)$, giving the number of cores in uniformly distributed but variably sized logarithmic bins of mass\textsuperscript{\bf \cite{MAU05}}. The blue histogram in Fig.~2b displays the cumulative CMF of W43-MM1, $N(>\!\log(M))$, giving the number of cores with $M_{\rm core}$ larger than $M$. If these histograms are truncated below the 90\% completeness limit ($\sim\!1.6\,\msun$), thereby reducing the sample to 105 cores, they can both be fitted with power-laws: $dN/d\log(M)\propto M^{\gamma_{\rm diff}}$ with $\gamma_{\rm diff}\simeq -0.90\pm0.06$, and $N(>\!\log(M))\propto M^{\gamma_{\rm cumul}}$ with $\gamma_{\rm cumul}\simeq -0.96\pm0.02$ (red lines on 2a and 2b). For comparison, the magenta lines on Figs. 2a and 2b show the slope of the IMF at masses larger than $\sim\!1\,\msun$, which for both the differential and cumulative forms of the IMF is $-1.35$\textsuperscript{\bf \cite{salpeter55,kroupa13}}.The range of core masses from $\sim\!1.6\,\msun$ to $\sim\!100\,\msun$ corresponds to a range of stellar masses smaller by a factor $1/\epsilon\sim2.5$ (i.e. stellar masses from $\sim\!0.6\,\msun$ to $\sim\!40\,\msun$), allowing a robust comparison with the higher-mass ($\stackrel{>}{\sim}\msun$) IMF. The result is stable against variations in the temperature model, dust emissivity, extraction algorithm, and reduction technique (see Supplementary Table~2). We conclude that, at masses larger than $\sim\!1.6\,\msun$, the CMF in W43-MM1 is markedly flatter than the IMF. This result seriously challenges the widespread assumption that the shape of the IMF is inherited directly from the CMF.

The main uncertainty on the CMF derives from the estimation of core masses, using their measured $1.3\,{\rm mm}$ continuum fluxes (see Methods). For the most massive cores, we use a $1.3\,{\rm mm}$ image based on the signal in a narrow composite band ($\sim\!65\,{\rm MHz}$ wide) that is not contaminated by line emission. For the lower mass cores (the majority), we use an image based on the signal in the full band ($\sim\!1.9\,{\rm GHz}$ wide), since line contamination for low-mass cores is expected to be negligible (see $C_{\rm line}$ in Supplementary Table~1). Variations in dust emissivity are presumed to be small among the W43-MM1 cores, since 90\% of them have uniformly high mean densities (${\bar n}_{_{{\rm H}_2}}\!\sim\!3\times 10^7-3\times 10^8\,{\rm cm}^{-3}$) and warm temperatures (${\bar T}_{\rm dust}\!\sim\!23\pm 2\,{\rm K}$) (see Supplementary Table~1). The main source of uncertainty is the dust temperature, since this is critical for converting flux into mass. Fig.~3 displays the mean dust temperatures used for estimating core masses. A mean line-of-sight dust temperature is estimated using \emph{Herschel} $70\;{\rm to}\;500\,\mu{\rm m}$ images\textsuperscript{\bf \cite{nguyen13}}, APEX $350\;{\rm and}\;870\,\mu{\rm m}$ images\textsuperscript{\bf \cite{nguyen11b}}, and mosaics obtained with the ALMA (present data) and IRAM\textsuperscript{\bf \cite{louvet14}} interferometers at $1.3\;{\rm and}\;3\,{\rm mm}$ respectively. By applying the Bayesian {\sc PPMAP} procedure\textsuperscript{\bf \cite{marsh15}}, we obtain column-density maps in different dust-temperature slices. The mean dust temperature in each pixel is then a column-density weighted average along that line of sight. However, in the vicinity of ten hot cores (of which three have been identified previously\textsuperscript{\bf \cite{herpin12,sridharan14}}), the local heating is not properly traced by the $2.5''$-resolution {\sc PPMAP} temperature image. Here, we divide the total luminosity of  the W43-MM1 cloud ($\sim\!2\times 10^4\,\lsun$) between the cores in proportion to their associated line contamination in the wide $1.3\,{\rm mm}$ band. We then use the individual luminosities, $L_\star\!<\!10\,\lsun$ up to $\sim\!10^4\,\lsun$ and {\sc PPMAP} background temperatures to estimate the core dust temperatures (${\bar T}_{\rm dust}\!\sim\!20\,{\rm K}$ to $\sim\!90\,{\rm K}$) and their uncertainties (see Supplementary Table~1), using an approximate radiation transport model\textsuperscript{\bf \cite{MA01}}. Monte-Carlo simulations indicate that the mass uncertainties in Supplementary Table~1 correspond to a $5\sigma$ uncertainty of $\pm 0.13$ in the slope of the CMF (see Methods and black-hatched area in Fig.~2b).

The fidelity of the CMF also depends on whether we have correctly identified cores, i.e. structures that (a) are gravitationally bound, and are therefore destined to spawn stars, and (b) already contain most of the mass that will eventually end up in those stars. Parenthetically, we note that core masses probably grow with time, due to inflowing gas streams like those observed towards many massive cores\textsuperscript{\bf \cite{HaFu08,SwWi08,csengeri14}}. In respect of criterion (a), studies of gravitational boundedness, using $^{13}$CS (5--4) lines from the present ALMA project to determine internal velocity dispersions, indicate that W43-MM1 cores with $M_{\rm core}>\!12\,\msun$ are secure, but the status of lower-mass cores would benefit from further investigation. In respect of criterion (b), the low luminosity-to-mass ratio of the whole region, $L_{\rm bol}/M\sim 5\,\lsun/\msun$\textsuperscript{\bf \cite{motte03}}, and the low mean temperature, ${\bar T}_{\rm dust}\sim 20\,{\rm K}$ (see Fig.~3) imply that the region is young. We note that the two most massive cores may assimilate further mass from their dense surroundings, and eventually form stars of  $\sim\!100\,\msun$, with $L_\star \!\sim\!10^5\,\lsun$ on the Main Sequence. However, currently they are $10-50$ times less luminous. We conclude that any protostars embedded within the W43-MM1 cores are at the very beginning of their accretion phase, and therefore contain only a small fraction of their final stellar mass.

Finally, the fidelity of the CMF depends on the completeness of the core sample. Due to increased source and background confusion in the denser parts of the cloud, the core sample is 90\% complete above $\sim\!1.6\,\msun$ outside the main filament, and above $\sim\!4.5\,\msun$ within it (see Supplementary Fig.~2). As in previous CMF evaluations, we use the median of the detection thresholds, here $\sim\!1.6\,\msun$, but even with the more conservative threshold of $\sim\!4.5\,\msun$, the CMF is still markedly shallower than the IMF (see Supplementary Table~2). The $5\sigma$ uncertainty in the fit to the CMF in Fig.~2b (black-hatched area) reflects the above assumptions, along with contributions from the data reduction method, the extraction algorithm\textsuperscript{\bf \cite{csengeri14}}, and the effects of source and background confusion (see Methods).

Due to the high sensitivity and resolution of our ALMA $1.3\,{\rm mm}$ image of the W43-MM1 cloud, we have been able to obtain, for the first time, a robust CMF covering masses in the range from $\sim\!1.6\,\msun$ to $\sim\!100\,\msun$, and hence to conclude that the CMF in W43-MM1 is much shallower than the higher-mass ($\stackrel{>}{\sim}\msun$) IMF, viz. $N(>\!\log(M))\propto M^{\gamma}$ with $\gamma\simeq -0.96\pm0.13$ instead of $\gamma\simeq -1.35$. This is in stark contrast with previous robust evaluations of the CMF in other regions and at lower masses, where the CMF appears to be very similar to the IMF\textsuperscript{\bf \cite{motte98, TeSa98, enoch08, konyves15}}. Scenarios that might explain our result and still allow the shape of the IMF to be inherited from the shape of the CMF, with an approximately self-similar mapping, fall into two broad categories. [A] W43-MM1 is not representative of the environments in which most low-mass stars form\textsuperscript{\bf \cite{harayama08,maia16}}; to obtain the complete IMF, stars formed in environments like W43-MM1 must be mixed with stars formed in environments, which contain a higher proportion of low-mass cores. However, this scenario is hard to justify, if the local IMF observed in extreme star-formation regions is similar to the universal IMF\textsuperscript{\bf \cite{bastian10}}. [B] Massive cores are over-represented in W43-MM1. {\it Either} [B1] massive cores live significantly longer than lower-mass cores\textsuperscript{\bf \cite{clark07}}, but current lifetime derivations\textsuperscript{\bf \cite{motte18}} suggest the opposite. {\it Or} [B2] star formation in regions like W43-MM1 is prolonged, and massive cores and the stars they spawn only form during a short period (probably near the beginning), whilst lower-mass cores and low-mass stars form over a much longer period; this seems unlikely, because it would require the low-mass period to be nearly ten times longer than the high-mass period, but it cannot be ruled out. If none of the above scenarios obtains, our results imply that the mapping from the CMF to the IMF is not statistically self-similar. Either higher-mass cores must convert a smaller fraction of their mass into stars than lower-mass cores (which seems unlikely\textsuperscript{\bf \cite{bontemps10}}); or they must spawn more stars, with a wider logarithmic range of masses (which seems more likely); or some combination of these possibilities. The shape of the IMF, at least at masses greater than $\sim\!\msun$, is then not simply inherited from the shape of the CMF and the processes that determine the IMF remain to be determined.

\subsection{Online Content.} Methods, along with Supplementary display items, are available in the online version of the paper. References unique to these sections appear only in the online paper.

\subsection{References}~\\

\begin{addendum}
\item [Acknowledgements] This paper makes use of the following ALMA data: \#2013.1.01365.S. ALMA is a partnership of ESO (representing its member states), NSF (USA) and NINS (Japan), together with NRC (Canada), MOST and ASIAA (Taiwan), and KASI (Republic of Korea), in cooperation with the Republic of Chile. The Joint ALMA Observatory is operated by ESO, AUI/NRAO and NAOJ. This project has received funding from the European Union's Horizon 2020 research and innovation programme StarFormMapper under grant agreement No 687528. This work was supported by the Programme National de Physique Stellaire and Physique et Chimie du Milieu Interstellaire (PNPS and PCMI) of CNRS/INSU (with INC/INP/IN2P3) co-funded by CEA and CNES. A.P.W. gratefully acknowledges the support of a consolidated grant (ST/K00926/1) from the UK Science and Technology Funding Council. T.Cs. acknowledges support from the Deut\-sche For\-schungs\-ge\-mein\-schaft, DFG  via the SPP (priority programme) 1573 `Physics of the ISM'. A.J.M. has received funding from the European Research Council (ERC) under the European Union's Horizon 2020 research and innovation programme (MagneticYSOs, grant agreement No 679937).
\item[Author Contributions] F.M. and F.L. led the project. E.C., T.N., F.M., and A.J.M. reduced the ALMA data. F.L. ran \emph{getsources} and the CASA simulator, T.Cs. ran {\sc MRE-GaussClumps}, and K.A.M. ran {\sc PPMAP}. S.B. and A.M. performed the Monte-Carlo simulations. F.M., T.N. and F.L. analysed CMF results. F.M and A.P.W. wrote the manuscript. F.M, S.B., F.L., Q.N.L., A.J.M. and P.S. contributed to the ALMA proposal. All authors discussed the results and implications, and commented on the manuscript.
\item[Competing Interests] The authors declare that they have no competing financial interests.
\item[Author Information] Reprints and permission information is available at www.nature.com/reprints. Readers are welcome to comment on the online version of the paper. Correspondence and requests for materials should be addressed to the corresponding authors, F.M. (frederique.motte@univ-grenoble-alpes.fr), T.N. (thomas.nony@univ-grenoble-alpes.fr), or F.L. (fabien.louvet@gmail.com).
\end{addendum}


\begin{center}
\begin{figure}
\centerline{\includegraphics[width=22cm,angle=0]{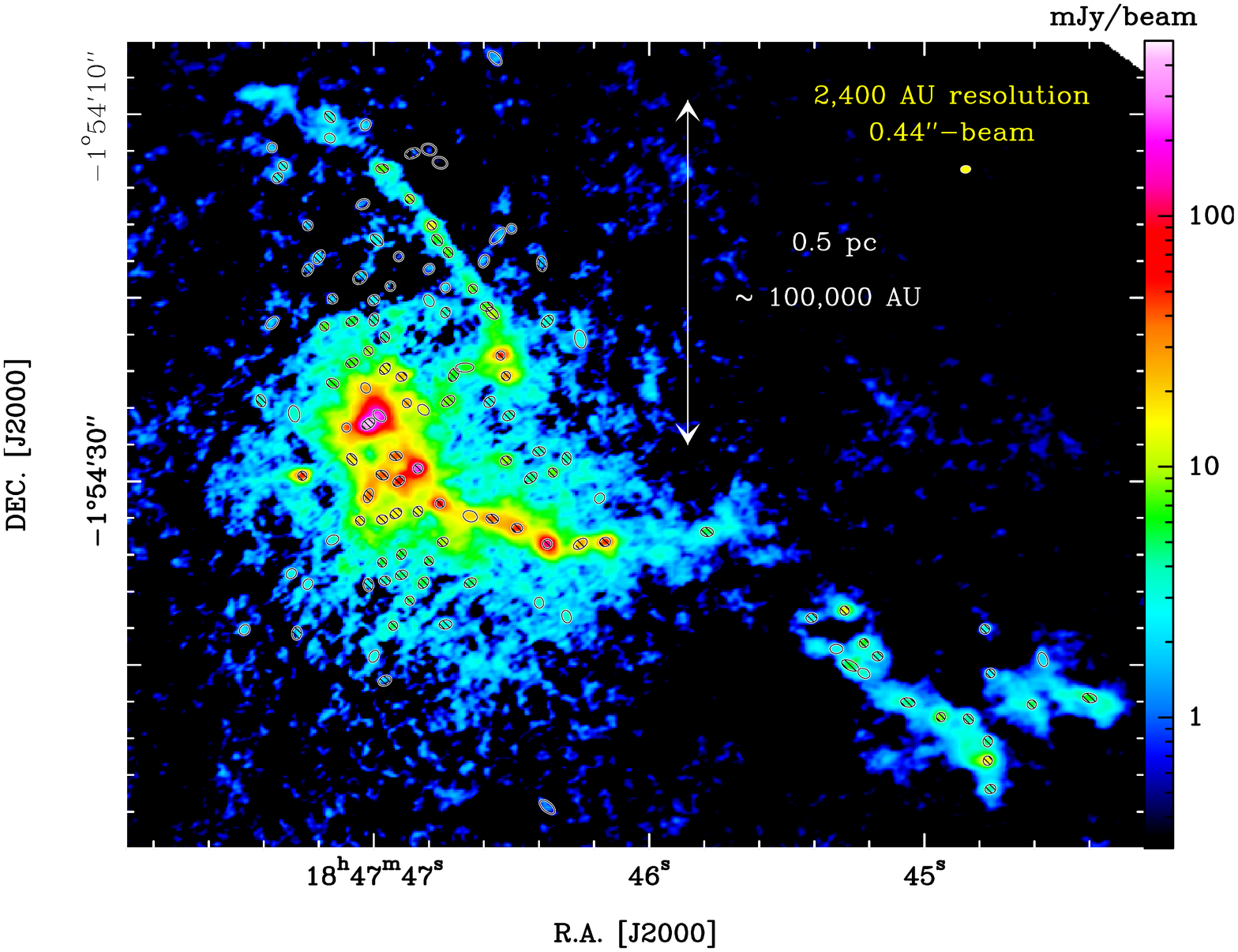}}
\vskip  -1.7cm
\caption{\textbf{High-angular resolution image of the W43-MM1 cloud, revealing a rich population of cores.} 
$1.3\,{\rm mm}$ dust continuum emission, observed by the ALMA interferometer, is presumed to trace the column density of gas, revealing high-density filaments and embedded cores. The filled yellow ellipse on the right represents the angular resolution, and a scale bar is shown. Ellipses outline core boundaries (at half-maximum) as defined by the {\it getsources}\textsuperscript{\bf 
\cite{menshchikov12}} 
extraction algorithm. Core masses span the range from $\sim\!1\,\msun$ to $\sim\!100\,\msun$, and can therefore be expected to spawn stars with masses from $\sim\!0.4\,\msun$ to $>\!40\,\msun$\textsuperscript{\bf 
 \cite{konyves15}} 
(see Supplementary Table~1). All cores are shown; hashed ellipses indicate the most robust identifications.}
\end{figure}
\end{center}

\begin{figure}
\vskip -2cm
\includegraphics[width=14.8cm]{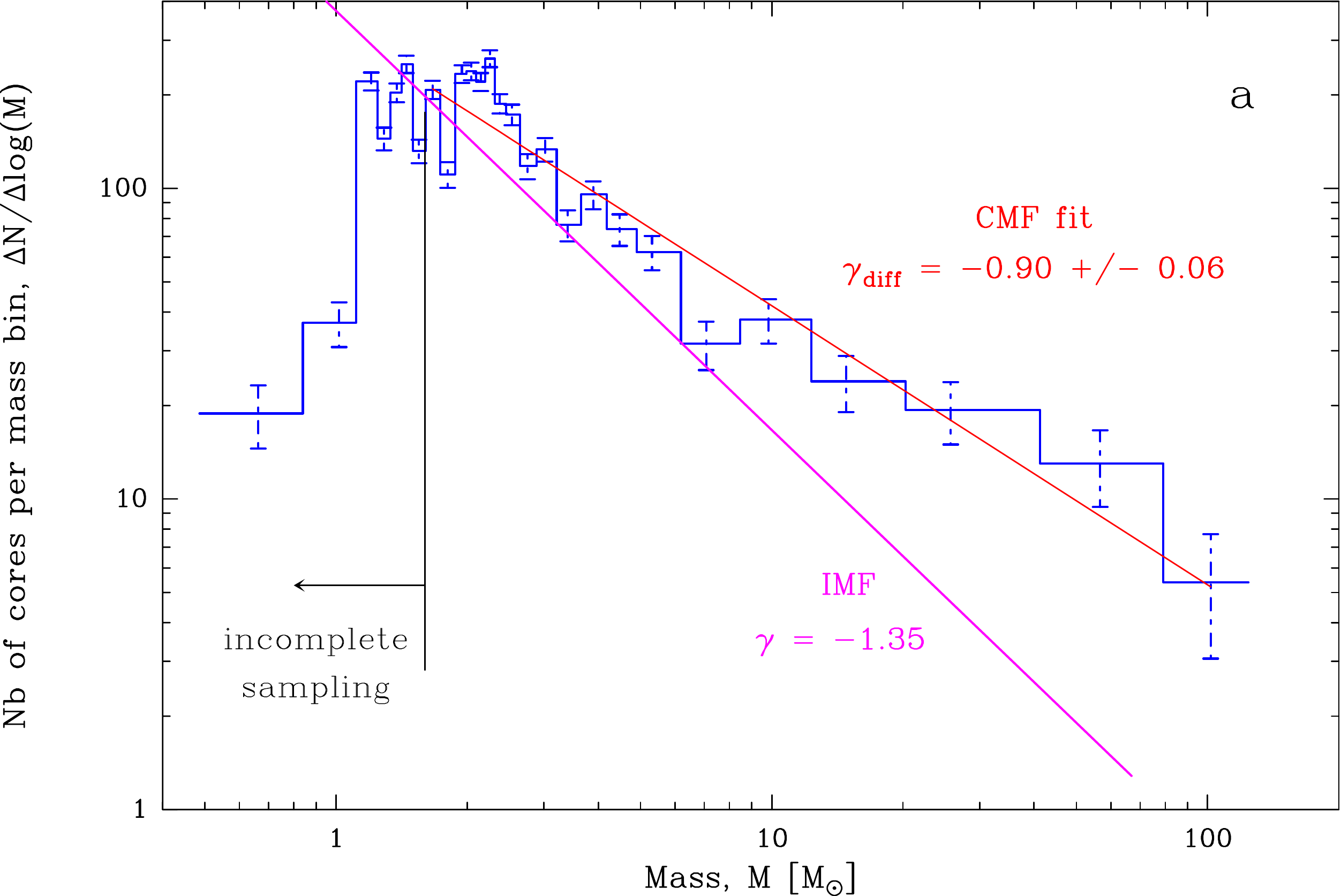}
\vskip -1.8cm
\includegraphics[width=18cm]{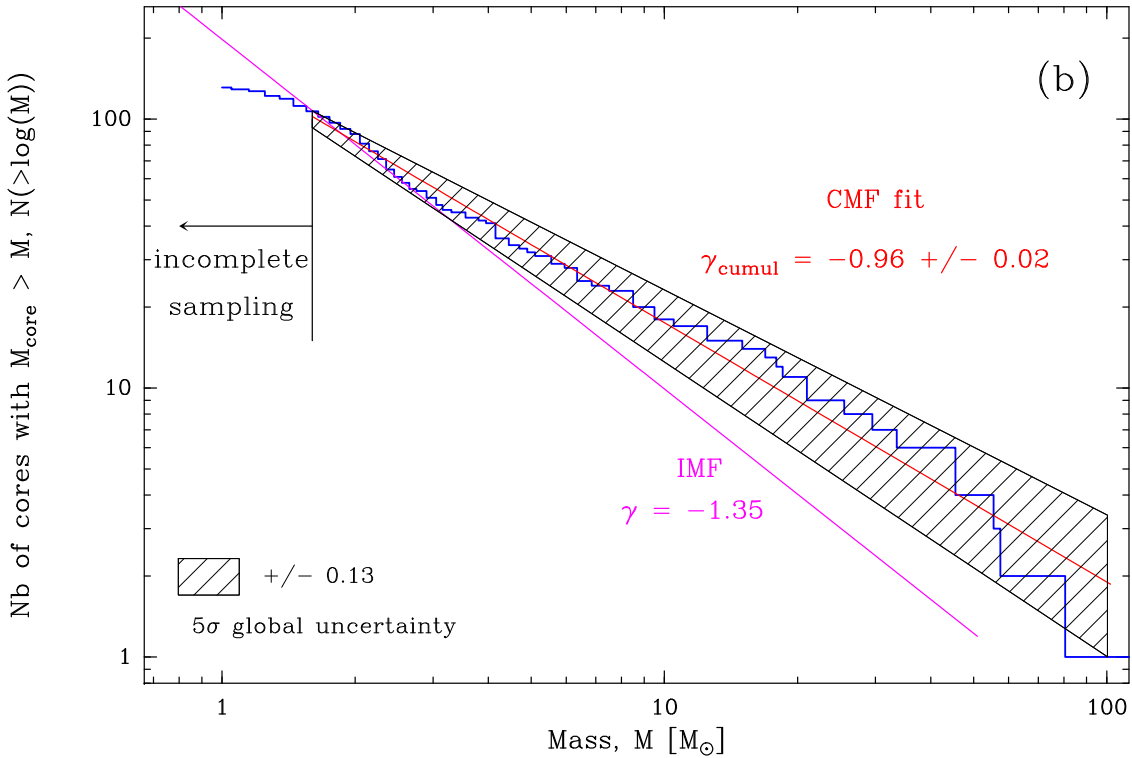}
\vskip -1.8cm
\caption{\textbf{The W43-MM1 core mass functions: (a) differential form; (b) cumulative form, challenging the relation between the CMF and the IMF.} 
Above the sample 90\% completeness limit, estimated to be $M_{\rm core} = 1.6\,\msun$ (black vertical line), the W43-MM1 CMFs (blue histograms) are well fitted by single power-laws: \textbf{(a)} $dN/d\log(M) \propto M^{-0.90}$, and \textbf{(b)} $N(>\!\log(M)) \propto M^{-0.96}$ (red lines and $1\sigma$ uncertainties). The error bars on the differential CMF correspond to $\sqrt N$ counting statistics. The cumulative CMF in \textbf{(b)} is the more robust, statistically; its $5\sigma$ global uncertainty ($\pm 0.13$, hatched area) is estimated from Monte-Carlo simulations. The W43-MM1 CMF is clearly flatter than the IMF\textsuperscript{\bf 
\cite{salpeter55,kroupa13}}, 
which in the corresponding mass range has slopes $dN/d \log(M) \propto M^{-1.35}$ and $N(>\!\log(M)) \propto M^{-1.35}$ (magenta lines).}
\end{figure}

\begin{figure}
\includegraphics[width=18cm]{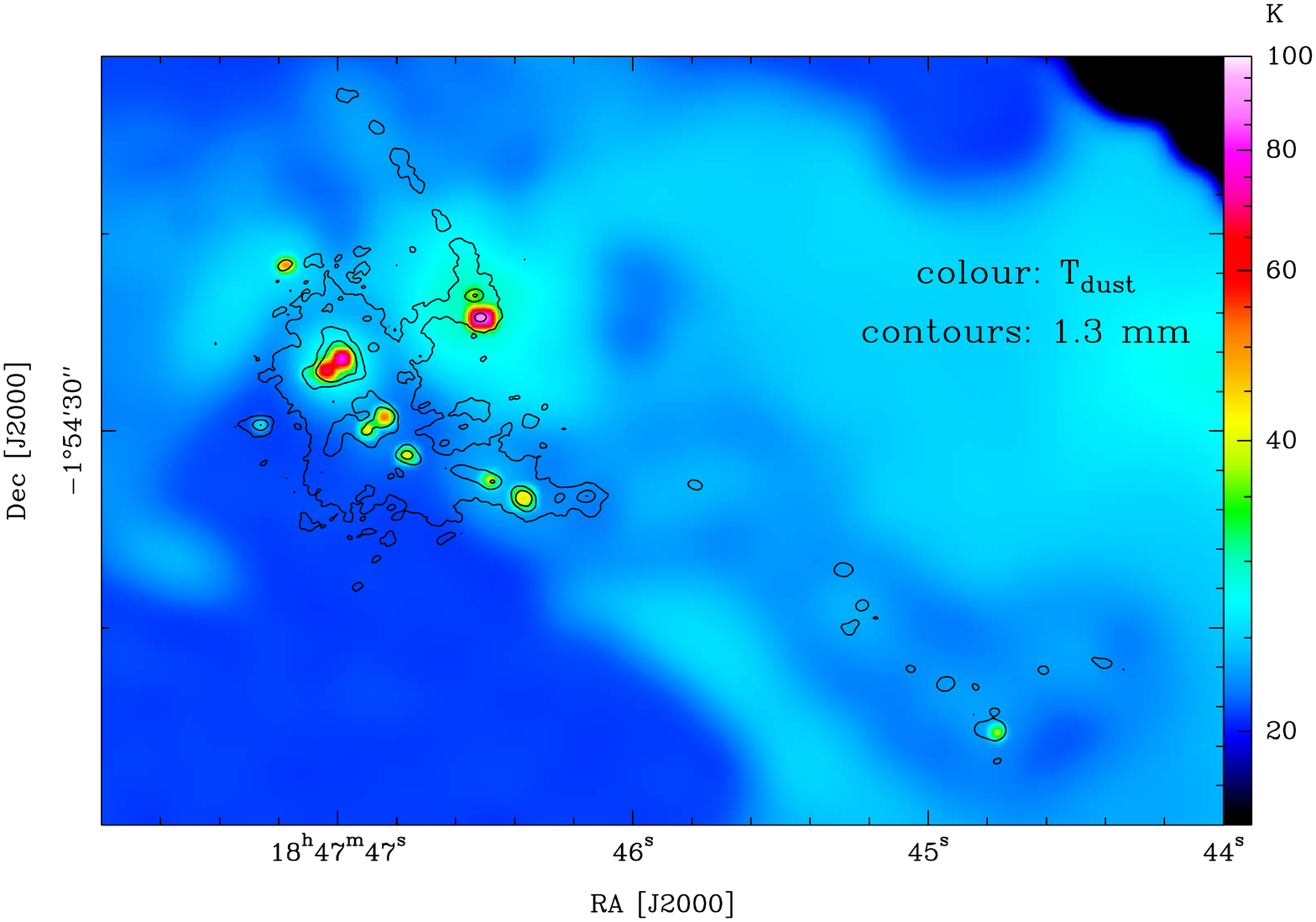}
\caption{\textbf{Mean estimated dust temperatures for cores}, ${\bar T}_{\rm dust}$. 
Over most of the frame, ${\bar T}_{\rm dust}$ is the column-density-weighted mean obtained with {\sc PPMAP}\textsuperscript{\bf 
\cite{marsh15}} 
using a $2.5''$ (or $14,000\,{\rm AU}$) resolution (colour scale). In the vicinity of the ten most luminous sources, the mean core temperature with $0.44''$ (or $4,200\,{\rm AU}$) resolution is calculated using an approximate radiation transport model based on their estimated luminosities\textsuperscript{\bf 
\cite{MA01}}. 
The W43-MM1 filaments are located by the 4, 20, and 50\,mJy/beam contours of the 1.3\,mm continuum map of Fig.~1. The ${\bar T}_{\rm dust}$ map shows cloud heating by the nearby cluster, screening by the W43-MM1 high-density filaments, and strong local heating by the ten most luminous ($30\;{\rm to}\;10^4~\lsun$) cores.}
\end{figure}

\newpage
\begin{methods}

\subsection{Observations and data reduction}

The W43-MM1 ALMA observations were carried out between September 2014 and June 2015 (project \#2013.1.01365.S) using the $12\,{\rm m}$ interferometric array. With baselines ranging from $13\,{\rm m}$ to $1045\,{\rm m}$, the $1.3\,{\rm mm}$ image is sensitive to emission on angular scales from the synthesised beam, $0.37''\!\times\!0.53''$ ($\sim\!2,400\,{\rm AU}\;{\rm at}\;5.5\,{\rm kpc}$), to the filtering scale of the interferometer for our observations, $12''$. The W43-MM1 cloud was imaged with a 33-field mosaic covering its $2.1\,{\rm pc}\!\times\!1.4\,{\rm pc}$ extent. The complete dataset contains eight spectral windows, of which the one most used here is dedicated to the $1.3\,{\rm mm}$ continuum centred at $233.45\,{\rm GHz}$, with $1.9\,{\rm GHz}$ bandwidth and $977\,{\rm kHz}$ channel step. The complementary data taken with the $7\,{\rm m}$ interferometric array were not used because of their low signal-to-noise ratio.

Data were reduced with the CASA~4.3.1 software\textsuperscript{\bf (31)},
applying manual and {\sc SelfCalibration} scripts. The {\sc SelfCalibration} technique improved image quality provided that it started from the highest-sensitivity map integrated over the complete 1.9~GHz continuum band. We then applied the {\sc Clean} algorithm with a robust weight of 0.5 and the {\sc MultiScale} option\textsuperscript{\bf (32)},
which minimises interferometric artefacts associated with missing short spacings. We did not use merged 12~m and 7~m data, because the {\sc MultiScale} cleaning of 12~m-only data provides the optimum image sensitivity for extracting cores. From the residual map of the continuum map integrated over the complete 1.9~GHz band, we estimate noise levels from $\sim$0.13~mJy/beam in the outskirts to $\sim$2.5~mJy/beam towards the main filament. This variation is due to confusion caused by strong, extended and structured cloud emission\textsuperscript{\bf (33)}.

\subsection{Core extraction}

Compact sources were extracted using \emph{getsources} (v1.140127), a multi-scale, multi-wavelength source-extraction algorithm developed for \emph{Herschel} CMF studies\textsuperscript{\bf (20; 34)}.
The main advantage of this algorithm applied to interferometric images is that it can extract sources in varying backgrounds, from strong filaments to negative areas associated with missing short spacings. We also used the multi-wavelength design of \emph{getsources} to simultaneously treat the highest-sensitivity ($1.9\,{\rm GHz}$-integrated) and line-free ($65\,{\rm MHz}$-integrated) $1.3\,{\rm mm}$ continuum images. During the detection step \emph{getsources} combines both images and defines a catalogue of sources with unique positions. At the measurement stage, \emph{getsources} measures the background and flux of these sources, on each map independently. 

We post-processed the \emph{getsources} catalogue to remove sources that are too extended, or whose ellipticity is too large to correspond to cores, or that are not centrally-peaked, through visual inspection. The final catalogue contains 131 sources with ${\rm Sig}\!>\!5$ (reliable), of which 94 cores have ${\rm Sig}\!>\!7$ (robust). Core characteristics are measured in the highest-sensitivity continuum image, except when core fluxes in the highest-sensitivity map are larger than those measured in the line-free image, due to line contamination. Sources with line contamination are massive cores harbouring hot cores. Supplementary Table~1 lists the cores detected in W43-MM1, and provides their significance level (Sig), coordinates, diameters, line contamination ratio, and peak and total integrated flux (corrected for line contamination and local background). While the Sig variable reflects the detection significance (equivalent to the signal-to-noise ratio of a source on the scale where it is best detected), flux uncertainties reflect the quality of the flux measurements.

\subsection{Core mass estimates}

The total mass of a core (gas plus dust), having uniform opacity throughout its solid angle, is given by
\begin{eqnarray}\nonumber
M_{\rm core} &=& -\, \frac{\Omega_{\rm beam} \;d^2} {\kappa_{1.3{\rm mm}}}\, 
	\ln\left(1\,-\,\frac{S^{\rm peak}_{1.3{\rm mm}}}{\Omega_{\rm beam}\;B_{1.3{\rm mm}}(T_{\rm dust})}\right)\,
	\times \frac{S^{\rm int}_{1.3{\rm mm}}} {S^{\rm peak}_{1.3{\rm mm}}} \\
	\nonumber
	&=& -\, M_{\rm core}^{\rm opt\;thin} \times
	\frac{\Omega_{\rm beam}\;B_{1.3{\rm mm}}(T_{\rm dust})}{S^{\rm peak}_{1.3{\rm mm}}}\,
	\ln\left(1\,-\,\frac{S^{\rm peak}_{1.3{\rm mm}}}{\Omega_{\rm beam}\;B_{1.3{\rm mm}}(T_{\rm dust})}\right)\,.
\end{eqnarray}
Here $\Omega_{\rm beam}$ is the solid angle of the beam, $d=5.5\,{\rm kpc}$ is the distance to the core, $\kappa_{\rm 1.3 \,mm}$ is the dust opacity per unit mass of gas and dust at $1.3\,{\rm mm}$, $S^{\rm peak}_{1.3{\rm mm}}$ and $S^{\rm int}_{1.3{\rm mm}}$ are the peak and integrated monochromatic fluxes of the core at $1.3\,{\rm mm}$, and $B_{\rm 1.3\,mm}(T_{\rm dust})$ is the Planck function at dust temperature $T_{\rm dust}$. $M_{\rm core}^{\rm opt\;thin}=S^{\rm int}_{1.3{\rm mm}}\;d^2/(\kappa_{1.3{\rm mm}}\;B_{1.3{\rm mm}}(T_{\rm dust}))$ is the core mass in the optically thin limit. We have adopted $\kappa_{\rm 1.3\,mm} = 0.01\,{\rm cm}^2\,{\rm g}^{-1}$\textsuperscript{\bf (35)}
as being most appropriate for high-density, cool to warm cores. Supplementary Table~1 lists for each core its mass and mean number density\textsuperscript{\bf (36)}.

The $1.3\,{\rm mm}$ continuum flux of a core arises mostly from thermal dust emission, which is optically thin except towards the seven densest, most massive cores. Contamination by free-free emission is estimated to be low ($<\!20\%$\textsuperscript{\bf \cite{motte03}}) for W43-MM1 cores, since none of the embedded protostars are sufficiently hot and luminous ($<10^3~\lsun$ for all except three cores) to ionise significant H{\sc ii} regions. Line contamination was evaluated by comparing the fluxes measured in the total $1.9\,{\rm GHz}$ continuum band and in a selection of line-free channels summing up to $65\,{\rm MHz}$. The mass uncertainties are derived from the flux uncertainties measured by \emph{getsources}, and from the estimated dust temperature error. We added the $\pm1$~K errors of {\sc PPMAP} temperatures\textsuperscript{\bf \cite{marsh15}} 
to the uncertainty derived from the self-shielding of $0.5''$ starless cores and/or the internal heating of $0.5''$ protostellar cores lying within the $2.5''$ PPMAP resolution\textsuperscript{\bf (26; 37)}.
We estimate the absolute values of the core masses to be uncertain by a factor of a few, and the relative values between cores to be uncertain by $\sim\!50\%$.

To estimate the sample completeness, we have performed ten Monte-Carlo simulations that place 2,000 synthetic cores, having masses in the range 1 to $10\,\msun$, on the worst-sensitivity background image determined by \emph{getsources} or on the original image containing real cores. Like the observed cores, the synthetic cores have small sizes, and this reduces confusion by nearby sources and gives an excellent measurement of the total core mass. We then simulated observation of these synthetic protoclusters and applied the same core extraction process as for the real data. Supplementary Fig.~2 shows that the $90\%$ completeness level depends weakly on the background intensity. It is $\sim\!0.8\pm0.1\,\msun$ in the outskirts of the protocluster, $\sim\!1.6\pm0.1\,\msun$ where most cores lie, and up to $\sim\!4.5\pm1\,\msun$ on the main filament (see the areas outlined in Supplementary Fig.~1). The $90\%$ completeness level also depends weakly on the source confusion, with an increase up to $\sim\!2.1\pm0.3\,\msun$ for overlapping sources, relative to the mean value, which remains at $\sim\!1.6\,\msun$  (see Supplementary Fig.~2).

\subsection{Core mass functions}

We have made numerous tests and Monte-Carlo simulations to prove the robustness of the CMF shape against interferometric artefacts, extraction methods (\emph{getsources}\textsuperscript{\bf \cite{menshchikov12}} or {\sc MRE-GaussClumps}\textsuperscript{\bf \cite{csengeri14}}), mass estimates, and CMF representations. In developing the best strategy to reduce the W43-MM1 ALMA image, the core catalogue steadily improved, with progressively fewer false cores and more solar-type cores. The 94 robust cores ($>\!7\sigma$) were detected in almost all runs. Supplementary Table~2 lists the major tests performed to evaluate the uncertainty of CMF fit. For statistical reasons, fits are more robust in cumulative form\textsuperscript{\bf (38)},
with the complete sample of 131 reliable cores, and the complete $1.6\,\msun$ to $100\,\msun$ mass range. However, Supplementary Table~2 shows that the higher-mass part ($>\!4.5\,\msun$) of the W43-MM1 CMF is still always flatter than the IMF. Monte-Carlo simulations (with 100,000 runs) show that the mass uncertainties correspond to the slope of the CMF, $\gamma_{\rm cumul}$, having a $5\sigma$ uncertainty of $\pm 0.13$. Building separate mass functions for cores that do, and cores that do not, contain accreting protostars is not feasible here. First, the most massive cores all contain young low-luminosity accreting protostars. Second, their outflows make it very hard to ascertain the precise nature of their lower-mass neighbours (i.e. whether or not they too already contain accreting protostars).

\subsection{Self-similar mapping}

A statistically self-similar mapping means that, for example, the probability that a core with mass in the range $1.0\,\msun$ to $1.1\,\msun$ spawns -- possibly along with other stars -- a star with mass in the range $0.40\,\msun$ to $0.44\,\msun$ is the same as the probability that a core with mass in the range $10\,\msun$ to $11\,\msun$ spawns a star with mass in the range $4.0\,\msun$ to $4.4\,\msun$. A mapping with small variance is required, because, if the variance were large, the peak of the IMF would be noticeably broader than the peak of the CMF, and this does not appear to be the case\textsuperscript{\bf \cite{konyves15}}. (We note that incompleteness at low core masses can only make the peak of the CMF broader than current estimates, and thereby strengthen this constraint.)

To compute a notional conversion efficiency (the mean of the total mass of stars subsequently spawned by a core divided by the core's mass), we need to determine: (i) the mean factor by which a core's mass grows, between when it is measured to determine the CMF, and when it has finished forming stars, $\mu_{\rm map}\;(>\!1)$; and (ii) the mean fraction of this final mass that goes into stars, $\eta_{\rm map}\;(<\!1)$. The notional efficiency is then $\mu_{\rm map}\eta_{\rm map}$, and can in principle exceed unity (which is why it is `notional'). If we also know (iii) the mean number of stars spawned by a single core, ${\cal N}_{\rm map}$, we can compute the upward shift from the peak of the IMF to the peak of the CMF, $1/\epsilon = {\cal N}_{\rm map}/ (\mu_{\rm map}\eta_{\rm map})$. Finally, if we know (iv) the logarithmic variance of the mean distribution of stellar masses spawned by a single core, $\sigma_{\rm map}$, this must be convolved with the logarithmic variance of the CMF to determine the logarithmic variance of the IMF, $\sigma_{_{\rm IMF}}^2=\sigma_{_{\rm CMF}}^2+\sigma_{\rm map}^2$. Since $\sigma_{\rm map}$ is necessarily finite, a self-similar mapping requires $\sigma_{_{\rm IMF}}\!>\!\sigma_{_{\rm CMF}}$.

\begin{addendum}
\item[Data Availability] 
The data that support the plots within this paper and other findings of this study are available from the corresponding authors (F.M., T.N., or F.L) upon reasonable request. Requests for materials should be addressed to F.M. (frederique.motte@univ-grenoble-alpes.fr), T.N. (thomas.nony@univ-grenoble-alpes.fr), or F.L. (fabien.louvet@gmail.com).
\end{addendum}

\section*{Additional references}
31. McMullin, J. P., Waters, B., Schiebel, D., Young, W. \& Golap, K. CASA Architecture and Applications. In Shaw, R. A., Hill, F. \& Bell, D. J. (eds.) 
{\it Astronomical Data Analysis Software and Systems XVI}, vol. 376 of {\it Astronomical Society of the Pacific Conference Series}, 127 (2007).\\\\
32. Rau, U. \& Cornwell, T. J. A multi-scale multi-frequency deconvolution algorithm for synthesis imaging in radio interferometry. {\it A\&A} {\bf 532}, A71 (2011).\\\\
33. Hennemann, M. {\it et al.} The spine of the swan: a Herschel study of the DR21 ridge and filaments in Cygnus X. {\it A\&A} {\bf 543}, L3 (2012).\\\\
34. Men'shchikov, A. A multi-scale filament extraction method: getfilaments. {\it A\&A} {\bf 560}, A63 (2013).\\\\
35. Ossenkopf, V. \& Henning, T. Dust opacities for protostellar cores. {\it A\&A} {\bf 291}, 943-959 (1994).\\\\
36. Tig\'e, J. {\it et al.} The earliest phases of high-mass star formation, as seen in NGC 6334 by Herschel-HOBYS. {\it A\&A} {\bf 602}, A77 (2017).\\\\
37. Bracco, A. {\it et al.} Probing changes of dust properties along a chain of solar-type prestellar and protostellar cores in Taurus with NIKA. {\it A\&A} {\bf 604}, A52 (2017).\\\\
38. Reid, M. A. \& Wilson, C. D. High-Mass Star Formation. III. The Functional Form of the Submillimeter Clump Mass Function. {\it ApJ} {\bf 650}, 970-984 (2006).

\end{methods}

\begin{figure}[htb!]
\vskip -1.7cm
\centerline{\includegraphics[width=21cm]{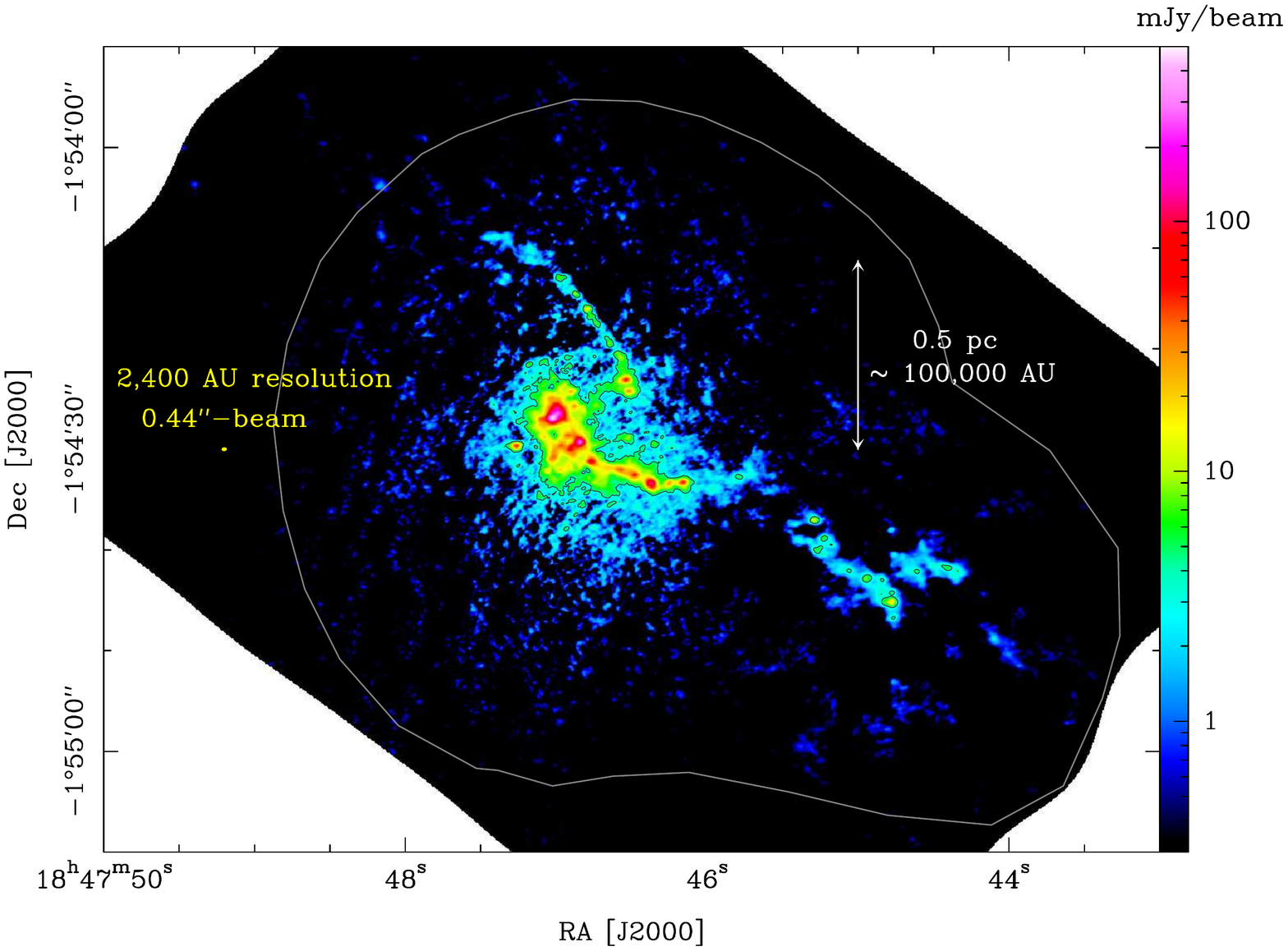}}
\vskip -2cm
\caption*{\textbf{Supplementary Figure 1 | High-angular resolution image of the complete W43-MM1 cloud.} 
$1.3\,{\rm mm}$ dust continuum emission, observed by the ALMA interferometer, is presumed to trace the column density of gas, revealing high-density filaments and embedded cores. The filled yellow ellipse on the left shows the angular resolution, and a scale bar is shown on the right. Contours correspond to the $S^{\rm peak}_{\rm 1.3mm}=4$\,mJy/beam flux level (black) and the $10^{23}$\,cm$^{-2}$ column density level from \emph{Herschel} (white). Despite the high-sensitivity of the ALMA image, and the high column densities observed, there appear to be no filaments or cores, with a completeness level of $\sim\!0.8\,\msun$, in the outskirts of the cloud. The core sample is 90\% complete down to $\sim\!1.6\,\msun$ on average, outside the main filament (between the two contours); and complete down to $\sim\!4.5\,\msun$ within it.}
\end{figure}

\begin{figure}[htb!]
\vskip -1.7cm
\centerline{\includegraphics[width=18cm]{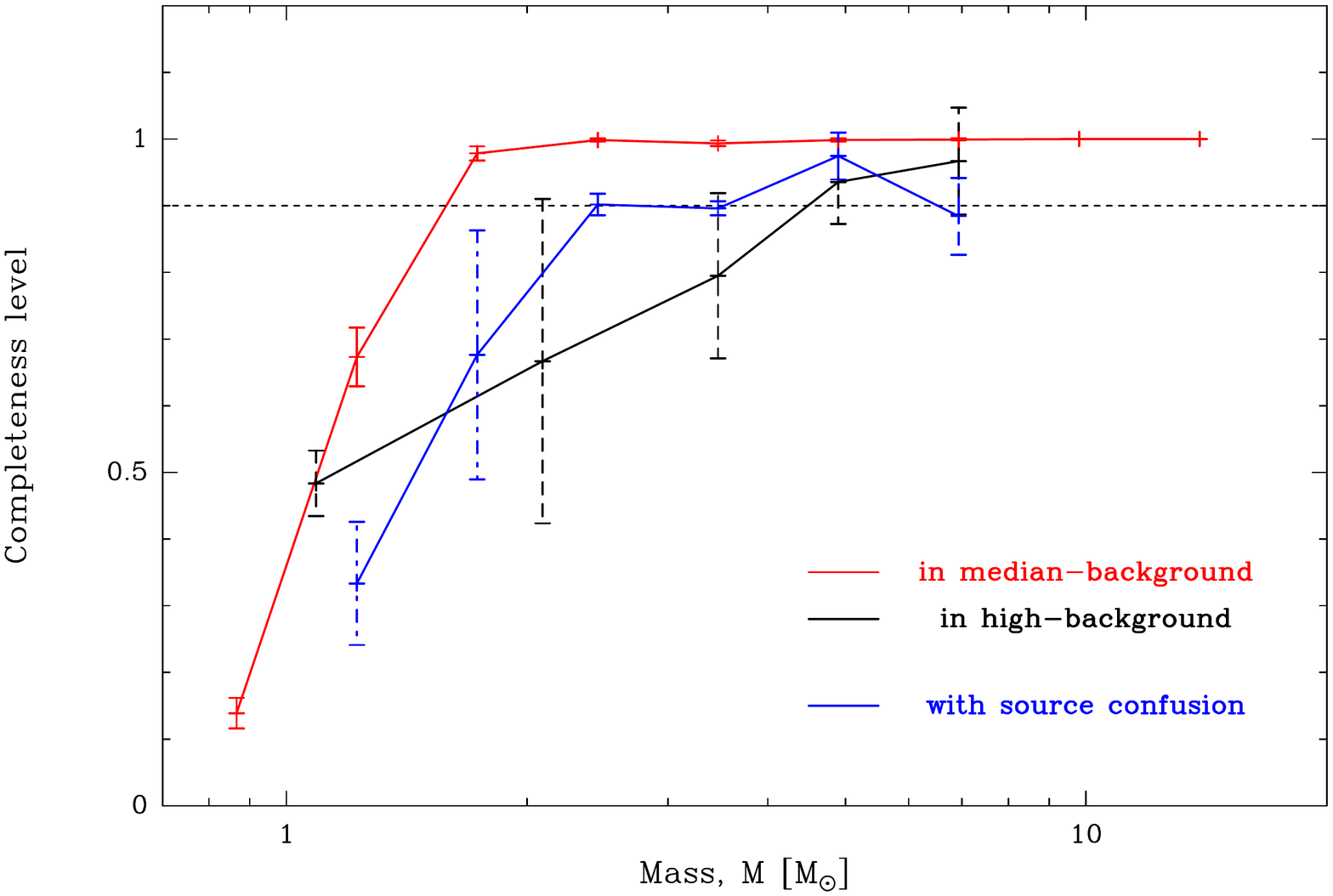}}
\vskip -1.5cm
\caption*{\textbf{Supplementary Figure 2 | Completeness level of the core sample of Supplementary Table~1.} 
We have performed Monte-Carlo simulations to investigate the sample completeness level as a function of background intensity (red and black curves) and source confusion (blue curve). Synthesized cores are placed (a) on the worst-sensitivity background image, which was determined by \emph{getsources} after subtracting cores of Supplementary Table~1 from Fig.~1, for background confusion tests; and (b) on the original image containing real cores, for source confusion tests. The observations are then simulated and the cores extracted as for the real data. Confusion, due to either a strong background, or the proximity of bright sources, is minimized by the multi-scale approach of \emph{getsources}. Simulations show that the core sample of Supplementary Table~1 is 90\% complete down to $\sim\!1.6 \pm0.1\,\msun$ (red curve) and to $\sim\!4.5\pm1\,\msun$ (black curve) in, respectively, the median- and high-background areas outlined in Supplementary Fig.~1. The census of cores overlapping with others remains 90\% complete down to $\sim\!2.4\pm0.2\,\msun$ (blue curve).}
\end{figure}

\setcounter{table}{0}
\begin{table}
	\small\footnotesize
   \caption*{}{\textbf{Supplementary Table 1} | Characteristics of the cores detected at 1.3\,mm in W43-MM1.}
   \makebox[\textwidth][c]{
\begin{tabular}{cccccccccccc} 
\hline    
\#    &       RA  &     Dec.   &      Size    &     FWHM        &      Sig     &  $C_{\rm line}$    &   $S^{\rm peak}_{\rm 1.3mm}$  &   $S^{\rm int}_{\rm 1.3mm}$    &   ${\bar T}_{\rm dust}$  &    $ M_{\rm core}$    &   ${\bar n}_{\rm H2}$\\
          &  [J2000]         &  [J2000]            &       [$'' \times ''$]      & [AU]    &       &   [$\%$]   &   [mJy/beam] &  [mJy]  &   [K]    &  [$\msun$] &  $ \times 10^7 \rm cm^{-3}$ \\
\hline    
1  &      18:47:47.02   &   -1:54:26.86   &     $0.82 \times 0.45$   &    2300   &     1100     &  45    &    $277 \pm 5$    &    $592 \pm 6$     &  $ 74 \pm 2$    &   $ 102 \pm 5$    &   $210 \pm 10$  \\
2  &      18:47:46.84   &   -1:54:29.30   &     $0.48 \times 0.45$   &    1200   &     670     &  26    &    $192 \pm 5$    &    $261 \pm 6$     &  $ 59 \pm 4$    &   $ 55 \pm 6$    &   $750 \pm 80$  \\
3  &      18:47:46.37   &   -1:54:33.41   &     $0.52 \times 0.47$   &    1200   &     330     &  13    &    $109 \pm 2$    &    $222 \pm 2$     &  $ 45 \pm 1$    &   $ 59 \pm 2$    &   $760 \pm 30$  \\
4  &      18:47:46.98   &   -1:54:26.42   &     $0.88 \times 0.45$   &    2500   &     340     &  74    &    $109 \pm 5$    &    $312 \pm 5$     &  $ 88 \pm 7$    &   $ 36 \pm 3$    &   $57 \pm 5$  \\
5  &      18:47:46.76   &   -1:54:31.21   &     $0.55 \times 0.45$   &    1300   &     170     &  15    &    $52 \pm 4$    &    $78 \pm 5$     &  $ 47 \pm 1$    &   $ 18 \pm 1$    &   $210 \pm 20$  \\
6  &      18:47:46.16   &   -1:54:33.30   &     $0.55 \times 0.45$   &    1300   &     130     &  0    &    $46.8 \pm 0.8$    &    $94 \pm 1$     &  $ 23 \pm 2$    &   $ 56 \pm 9$    &   $660 \pm 100$  \\
7  &      18:47:47.26   &   -1:54:29.70   &     $0.48 \times 0.45$   &    1200   &     130     &  0    &    $47 \pm 1$    &    $59 \pm 1$     &  $ 30 \pm 2$    &   $ 23 \pm 2$    &   $320 \pm 30$  \\
8  &      18:47:46.54   &   -1:54:23.15   &     $0.54 \times 0.45$   &    1200   &     130     &  0    &    $45 \pm 2$    &    $61 \pm 2$     &  $ 45 \pm 5$    &   $ 14 \pm 2$    &   $190 \pm 30$  \\
9  &      18:47:46.48   &   -1:54:32.54   &     $0.63 \times 0.45$   &    1600   &     110     &  17    &    $37 \pm 2$    &    $87 \pm 3$     &  $ 50 \pm 1$    &   $ 17.8 \pm 0.9$    &   $97 \pm 5$  \\
10  &      18:47:46.91   &   -1:54:29.99   &     $0.67 \times 0.5$   &    2100   &     91     &  24    &    $35 \pm 4$    &    $78 \pm 5$     &  $ 51 \pm 2$    &   $ 16 \pm 1$    &   $43 \pm 4$  \\
11  &      18:47:46.52   &   -1:54:24.26   &     $0.52 \times 0.45$   &    1200   &     72     &  60    &    $16 \pm 2$    &    $21 \pm 2$     &  $ 93 \pm 11$    &   $ 2.1 \pm 0.3$    &   $28 \pm 5$  \\
12  &      18:47:46.57   &   -1:54:32.04   &     $0.72 \times 0.45$   &    2000   &     77     &  0    &    $30 \pm 2$    &    $57 \pm 2$     &  $ 23 \pm 2$    &   $ 31 \pm 4$    &   $100 \pm 10$  \\
13  &      18:47:46.92   &   -1:54:28.62   &     $0.69 \times 0.45$   &    1900   &     63     &  0    &    $22 \pm 6$    &    $22 \pm 6$     &  $ 24 \pm 2$    &   $ 11 \pm 3$    &   $40 \pm 10$  \\
14  &      18:47:46.97   &   -1:54:29.66   &     $0.72 \times 0.45$   &    2000   &     57     &  0    &    $23 \pm 6$    &    $33 \pm 6$     &  $ 22 \pm 2$    &   $ 19 \pm 4$    &   $60 \pm 10$  \\
15  &      18:47:44.77   &   -1:54:45.22   &     $0.66 \times 0.54$   &    2200   &     62     &  54    &    $9.1 \pm 0.8$    &    $20 \pm 1$     &  $ 50 \pm 3$    &   $ 3.9 \pm 0.3$    &   $8.8 \pm 0.8$  \\
16  &      18:47:47.02   &   -1:54:30.78   &     $0.78 \times 0.45$   &    2200   &     68     &  0    &    $25 \pm 3$    &    $58 \pm 4$     &  $ 21 \pm 2$    &   $ 36 \pm 6$    &   $90 \pm 10$  \\
17  &      18:47:47.10   &   -1:54:27.07   &     $0.51 \times 0.45$   &    1200   &     46     &  0    &    $16 \pm 6$    &    $18 \pm 6$     &  $ 35 \pm 2$    &   $ 6 \pm 2$    &   $80 \pm 30$  \\
18  &      18:47:46.25   &   -1:54:33.41   &     $0.93 \times 0.45$   &    2600   &     59     &  0    &    $22.7 \pm 0.8$    &    $53 \pm 1$     &  $ 23 \pm 2$    &   $ 28 \pm 4$    &   $39 \pm 5$  \\
19  &      18:47:46.88   &   -1:54:25.74   &     $0.45 \times 0.45$   &    1200   &     45     &  0    &    $16 \pm 5$    &    $19 \pm 5$     &  $ 24 \pm 2$    &   $ 9 \pm 3$    &   $130 \pm 40$  \\
20  &      18:47:45.29   &   -1:54:37.04   &     $0.48 \times 0.45$   &    1200   &     47     &  0    &    $13.2 \pm 0.5$    &    $16.4 \pm 0.6$     &  $ 23 \pm 2$    &   $ 8 \pm 1$    &   $110 \pm 20$  \\
21  &      18:47:46.79   &   -1:54:16.06   &     $0.47 \times 0.45$   &    1200   &     46     &  0    &    $15 \pm 1$    &    $20 \pm 1$     &  $ 23 \pm 2$    &   $ 10 \pm 1$    &   $140 \pm 20$  \\
22  &      18:47:47.05   &   -1:54:32.15   &     $0.45 \times 0.45$   &    1200   &     40     &  0    &    $13 \pm 2$    &    $13 \pm 2$     &  $ 21 \pm 2$    &   $ 8 \pm 2$    &   $100 \pm 20$  \\
23  &      18:47:46.90   &   -1:54:24.30   &     $0.63 \times 0.45$   &    1600   &     44     &  0    &    $15 \pm 5$    &    $20 \pm 5$     &  $ 22 \pm 2$    &   $ 11 \pm 3$    &   $60 \pm 20$  \\
24  &      18:47:46.97   &   -1:54:32.08   &     $0.63 \times 0.45$   &    1600   &     29     &  0    &    $10 \pm 2$    &    $12 \pm 2$     &  $ 21 \pm 2$    &   $ 7 \pm 1$    &   $37 \pm 8$  \\
25  &      18:47:46.87   &   -1:54:14.62   &     $0.45 \times 0.45$   &    1200   &     29     &  0    &    $8 \pm 1$    &    $8 \pm 1$     &  $ 24 \pm 2$    &   $ 4.0 \pm 0.7$    &   $50 \pm 10$  \\
26  &      18:47:44.94   &   -1:54:42.84   &     $0.51 \times 0.45$   &    1200   &     28     &  0    &    $7.1 \pm 0.6$    &    $10.5 \pm 0.7$     &  $ 23 \pm 2$    &   $ 5.2 \pm 0.7$    &   $70 \pm 10$  \\
27  &      18:47:47.03   &   -1:54:24.91   &     $0.62 \times 0.45$   &    1600   &     25     &  0    &    $10 \pm 5$    &    $11 \pm 5$     &  $ 26 \pm 2$    &   $ 5 \pm 2$    &   $30 \pm 10$  \\
28  &      18:47:44.40   &   -1:54:41.80   &     $0.81 \times 0.45$   &    2300   &     22     &  0    &    $5.8 \pm 0.7$    &    $11.7 \pm 0.8$     &  $ 23 \pm 2$    &   $ 5.9 \pm 0.9$    &   $12 \pm 2$  \\
29  &      18:47:45.22   &   -1:54:38.81   &     $0.45 \times 0.45$   &    1200   &     20     &  0    &    $5.5 \pm 0.5$    &    $6.6 \pm 0.6$     &  $ 24 \pm 2$    &   $ 3.1 \pm 0.5$    &   $43 \pm 6$  \\
30  &      18:47:47.18   &   -1:54:21.56   &     $0.57 \times 0.45$   &    1400   &     19     &  26    &    $5 \pm 2$    &    $6 \pm 2$     &  $ 62 \pm 4$    &   $ 1.0 \pm 0.3$    &   $10 \pm 3$  \\
31  &      18:47:46.73   &   -1:54:17.53   &     $0.59 \times 0.45$   &    1500   &     19     &  0    &    $6.6 \pm 0.9$    &    $8 \pm 1$     &  $ 23 \pm 2$    &   $ 4.3 \pm 0.7$    &   $33 \pm 6$  \\
32  &      18:47:46.57   &   -1:54:20.84   &     $0.84 \times 0.45$   &    2400   &     20     &  0    &    $7 \pm 2$    &    $15 \pm 2$     &  $ 28 \pm 2$    &   $ 6 \pm 1$    &   $11 \pm 2$  \\
33  &      18:47:47.08   &   -1:54:28.80   &     $0.72 \times 0.46$   &    2000   &     24     &  0    &    $9 \pm 4$    &    $17 \pm 4$     &  $ 22 \pm 2$    &   $ 9 \pm 3$    &   $26 \pm 8$  \\
34  &      18:47:46.52   &   -1:54:28.87   &     $0.64 \times 0.45$   &    1700   &     18     &  0    &    $6.9 \pm 0.9$    &    $15 \pm 1$     &  $ 25 \pm 2$    &   $ 7 \pm 1$    &   $35 \pm 5$  \\
35  &      18:47:46.92   &   -1:54:31.75   &     $0.73 \times 0.46$   &    2100   &     20     &  0    &    $7 \pm 3$    &    $15 \pm 3$     &  $ 22 \pm 2$    &   $ 8 \pm 2$    &   $23 \pm 5$  \\
36  &      18:47:46.64   &   -1:54:19.51   &     $0.45 \times 0.45$   &    1200   &     18     &  0    &    $5.8 \pm 0.8$    &    $8 \pm 1$     &  $ 27 \pm 2$    &   $ 3.1 \pm 0.5$    &   $43 \pm 7$  \\
37  &      18:47:46.97   &   -1:54:12.96   &     $0.73 \times 0.45$   &    2000   &     20     &  0    &    $5 \pm 1$    &    $11 \pm 1$     &  $ 23 \pm 2$    &   $ 5.2 \pm 0.9$    &   $15 \pm 3$  \\
38  &      18:47:47.02   &   -1:54:22.90   &     $0.45 \times 0.45$   &    1200   &     16     &  0    &    $5 \pm 1$    &    $5 \pm 2$     &  $ 24 \pm 2$    &   $ 2.5 \pm 0.8$    &   $30 \pm 10$  \\
39  &      18:47:44.61   &   -1:54:42.16   &     $0.47 \times 0.45$   &    1200   &     16     &  0    &    $3.6 \pm 0.6$    &    $4.8 \pm 0.7$     &  $ 23 \pm 2$    &   $ 2.3 \pm 0.4$    &   $32 \pm 6$  \\
40  &      18:47:46.35   &   -1:54:29.52   &     $0.53 \times 0.45$   &    1200   &     17     &  0    &    $5.8 \pm 0.8$    &    $9.4 \pm 0.9$     &  $ 24 \pm 2$    &   $ 4.3 \pm 0.7$    &   $59 \pm 9$  \\
41  &      18:47:46.96   &   -1:54:23.87   &     $0.68 \times 0.45$   &    1800   &     15     &  0    &    $5 \pm 3$    &    $7 \pm 3$     &  $ 22 \pm 2$    &   $ 4 \pm 2$    &   $14 \pm 7$  \\
42  &      18:47:44.78   &   -1:54:38.05   &     $0.45 \times 0.45$   &    1200   &     13     &  0    &    $3.0 \pm 0.4$    &    $3.1 \pm 0.4$     &  $ 24 \pm 2$    &   $ 1.4 \pm 0.2$    &   $19 \pm 3$  \\
43  &      18:47:46.84   &   -1:54:31.64   &     $0.64 \times 0.46$   &    1700   &     12     &  0    &    $4 \pm 3$    &    $5 \pm 3$     &  $ 22 \pm 2$    &   $ 2 \pm 1$    &   $11 \pm 7$  \\
44  &      18:47:44.76   &   -1:54:46.76   &     $0.56 \times 0.45$   &    1300   &     15     &  0    &    $3.2 \pm 0.4$    &    $5.0 \pm 0.4$     &  $ 22 \pm 2$    &   $ 2.5 \pm 0.4$    &   $28 \pm 4$  \\
45  &      18:47:46.87   &   -1:54:36.50   &     $0.45 \times 0.45$   &    1200   &     11     &  0    &    $4.1 \pm 0.9$    &    $4.2 \pm 0.9$     &  $ 21 \pm 2$    &   $ 2.4 \pm 0.6$    &   $32 \pm 8$  \\
46  &      18:47:46.77   &   -1:54:16.85   &     $0.73 \times 0.45$   &    2000   &     12     &  0    &    $4.7 \pm 0.9$    &    $8.3 \pm 0.9$     &  $ 21 \pm 2$    &   $ 4.6 \pm 0.8$    &   $14 \pm 2$  \\
47  &      18:47:46.65   &   -1:54:31.90   &     $0.81 \times 0.59$   &    2900   &     21     &  0    &    $9 \pm 2$    &    $35 \pm 3$     &  $ 22 \pm 2$    &   $ 19 \pm 3$    &   $18 \pm 3$  \\
48  &      18:47:46.82   &   -1:54:26.10   &     $0.66 \times 0.45$   &    1700   &     12     &  0    &    $5 \pm 2$    &    $7 \pm 2$     &  $ 23 \pm 2$    &   $ 4 \pm 1$    &   $16 \pm 6$  \\
49  &      18:47:46.59   &   -1:54:20.48   &     $0.73 \times 0.48$   &    2200   &     14     &  0    &    $5 \pm 1$    &    $9 \pm 2$     &  $ 28 \pm 2$    &   $ 3.7 \pm 0.8$    &   $9 \pm 2$  \\
50  &      18:47:46.97   &   -1:54:34.42   &     $0.45 \times 0.45$   &    1200   &     11     &  0    &    $3 \pm 1$    &    $3 \pm 1$     &  $ 21 \pm 2$    &   $ 1.7 \pm 0.8$    &   $20 \pm 10$  \\
\hline   
\end{tabular} 
 	}
   \label{table}
   \end{table}

   \setcounter{table}{0}
   \begin{table}
      \small\footnotesize
   \caption*{}{\textbf{Supplementary Table 1} | Characteristics of the cores detected at 1.3\,mm in W43-MM1 (continued).}
   \makebox[\textwidth][c]{
\begin{tabular}{cccccccccccc}  
\hline
\#    &       RA  &     Dec.   &      Size    &     FWHM        &      Sig     &  $C_{\rm line}$    &   $S^{\rm peak}_{\rm 1.3mm}$  &   $S^{\rm int}_{\rm 1.3mm}$    &   ${\bar T}_{\rm dust}$  &    $ M_{\rm core}$    &   ${\bar n}_{\rm H2}$\\
          &  [J2000]         &  [J2000]            &       [$'' \times ''$]      & [AU]    &       &   [$\%$]   &   [mJy/beam] &  [mJy]  &   [K]    &  [$\msun$] &  $ \times 10^7 \rm cm^{-3}$ \\
\hline    
51  &      18:47:45.27   &   -1:54:40.03   &     $0.96 \times 0.52$   &    3000   &     15     &  0    &    $4.0 \pm 0.7$    &    $10.2 \pm 0.9$     &  $ 24 \pm 2$    &   $ 4.8 \pm 0.7$    &   $4.3 \pm 0.6$  \\
52  &      18:47:46.80   &   -1:54:34.34   &     $0.45 \times 0.45$   &    1200   &     10     &  0    &    $3 \pm 1$    &    $3 \pm 1$     &  $ 21 \pm 2$    &   $ 1.8 \pm 0.8$    &   $20 \pm 10$  \\
53  &      18:47:47.05   &   -1:54:18.90   &     $0.77 \times 0.55$   &    2600   &     11     &  0    &    $3.8 \pm 0.7$    &    $7.5 \pm 0.8$     &  $ 22 \pm 2$    &   $ 4.0 \pm 0.7$    &   $5.2 \pm 0.9$  \\
54  &      18:47:45.06   &   -1:54:42.05   &     $0.79 \times 0.45$   &    2200   &     11     &  0    &    $2.9 \pm 0.4$    &    $6.0 \pm 0.5$     &  $ 23 \pm 2$    &   $ 3.0 \pm 0.4$    &   $7 \pm 1$  \\
55  &      18:47:46.96   &   -1:54:35.42   &     $0.63 \times 0.45$   &    1600   &     9.6     &  0    &    $3 \pm 1$    &    $3 \pm 1$     &  $ 21 \pm 2$    &   $ 2.0 \pm 0.8$    &   $11 \pm 4$  \\
56  &      18:47:47.00   &   -1:54:20.12   &     $0.58 \times 0.45$   &    1400   &     11     &  0    &    $3 \pm 1$    &    $3 \pm 1$     &  $ 21 \pm 2$    &   $ 1.9 \pm 0.9$    &   $17 \pm 8$  \\
57  &      18:47:46.94   &   -1:54:19.37   &     $0.45 \times 0.45$   &    1200   &     9.8     &  0    &    $3 \pm 1$    &    $3 \pm 1$     &  $ 23 \pm 2$    &   $ 1.7 \pm 0.5$    &   $23 \pm 8$  \\
58  &      18:47:47.08   &   -1:54:21.28   &     $0.74 \times 0.45$   &    2000   &     11     &  0    &    $4 \pm 2$    &    $7 \pm 3$     &  $ 25 \pm 2$    &   $ 3 \pm 1$    &   $8 \pm 4$  \\
59  &      18:47:44.77   &   -1:54:44.17   &     $0.63 \times 0.45$   &    1600   &     9.9     &  0    &    $2.3 \pm 0.9$    &    $3 \pm 1$     &  $ 23 \pm 2$    &   $ 1.4 \pm 0.5$    &   $8 \pm 3$  \\
60  &      18:47:44.76   &   -1:54:40.46   &     $0.48 \times 0.45$   &    1200   &     10     &  0    &    $2.4 \pm 0.8$    &    $3.2 \pm 0.9$     &  $ 23 \pm 2$    &   $ 1.6 \pm 0.5$    &   $22 \pm 7$  \\
61  &      18:47:46.82   &   -1:54:35.53   &     $0.7 \times 0.53$   &    2300   &     12     &  0    &    $4 \pm 1$    &    $10 \pm 1$     &  $ 21 \pm 2$    &   $ 6 \pm 1$    &   $11 \pm 2$  \\
62  &      18:47:45.79   &   -1:54:32.76   &     $0.65 \times 0.45$   &    1700   &     11     &  0    &    $3.7 \pm 0.7$    &    $9 \pm 1$     &  $ 24 \pm 2$    &   $ 4.0 \pm 0.6$    &   $19 \pm 3$  \\
63  &      18:47:47.33   &   -1:54:12.82   &     $0.47 \times 0.45$   &    1200   &     9.5     &  0    &    $3.0 \pm 0.7$    &    $3.8 \pm 0.7$     &  $ 21 \pm 2$    &   $ 2.1 \pm 0.5$    &   $29 \pm 7$  \\
64  &      18:47:46.99   &   -1:54:16.85   &     $0.83 \times 0.49$   &    2600   &     10     &  0    &    $3 \pm 1$    &    $6 \pm 1$     &  $ 22 \pm 2$    &   $ 3.2 \pm 0.7$    &   $5 \pm 1$  \\
65  &      18:47:45.41   &   -1:54:37.44   &     $0.56 \times 0.45$   &    1300   &     8.9     &  0    &    $2.7 \pm 0.6$    &    $3.1 \pm 0.6$     &  $ 23 \pm 2$    &   $ 1.6 \pm 0.4$    &   $16 \pm 4$  \\
66  &      18:47:47.00   &   -1:54:21.20   &     $0.68 \times 0.45$   &    1800   &     9.0     &  0    &    $3 \pm 2$    &    $4 \pm 2$     &  $ 22 \pm 2$    &   $ 2.3 \pm 0.9$    &   $9 \pm 3$  \\
67  &      18:47:44.09   &   -1:54:48.89   &     $0.66 \times 0.47$   &    1900   &     9.8     &  0    &    $1.8 \pm 0.3$    &    $4.7 \pm 0.4$     &  $ 23 \pm 2$    &   $ 2.2 \pm 0.3$    &   $8 \pm 1$  \\
68  &      18:47:47.02   &   -1:54:35.64   &     $0.71 \times 0.45$   &    2000   &     8.7     &  0    &    $3 \pm 1$    &    $4 \pm 1$     &  $ 21 \pm 2$    &   $ 2.2 \pm 0.6$    &   $7 \pm 2$  \\
69  &      18:47:47.15   &   -1:54:20.05   &     $0.5 \times 0.45$   &    1200   &     9.1     &  0    &    $3 \pm 1$    &    $5 \pm 1$     &  $ 25 \pm 2$    &   $ 2.1 \pm 0.6$    &   $29 \pm 8$  \\
70  &      18:47:47.20   &   -1:54:17.78   &     $0.79 \times 0.48$   &    2400   &     9.1     &  0    &    $3.3 \pm 0.7$    &    $6.8 \pm 0.8$     &  $ 23 \pm 2$    &   $ 3.4 \pm 0.6$    &   $6 \pm 1$  \\
71  &      18:47:47.35   &   -1:54:13.46   &     $0.62 \times 0.45$   &    1600   &     8.3     &  0    &    $2.4 \pm 0.6$    &    $3.1 \pm 0.6$     &  $ 21 \pm 2$    &   $ 1.7 \pm 0.4$    &   $10 \pm 2$  \\
72  &      18:47:46.65   &   -1:54:35.53   &     $0.74 \times 0.45$   &    2100   &     8.2     &  0    &    $3 \pm 1$    &    $4 \pm 1$     &  $ 20 \pm 2$    &   $ 2.3 \pm 0.8$    &   $6 \pm 2$  \\
73  &      18:47:44.84   &   -1:54:42.95   &     $0.63 \times 0.53$   &    2100   &     11     &  0    &    $2.5 \pm 0.8$    &    $6 \pm 1$     &  $ 23 \pm 2$    &   $ 2.8 \pm 0.7$    &   $8 \pm 2$  \\
74  &      18:47:45.17   &   -1:54:39.53   &     $0.62 \times 0.45$   &    1600   &     8.0     &  0    &    $2.2 \pm 0.4$    &    $2.9 \pm 0.5$     &  $ 24 \pm 2$    &   $ 1.4 \pm 0.3$    &   $8 \pm 2$  \\
75  &      18:47:46.90   &   -1:54:35.10   &     $0.73 \times 0.45$   &    2000   &     8.0     &  0    &    $2 \pm 1$    &    $3 \pm 1$     &  $ 21 \pm 2$    &   $ 1.9 \pm 0.7$    &   $6 \pm 2$  \\
76  &      18:47:46.93   &   -1:54:37.87   &     $0.48 \times 0.45$   &    1200   &     8.0     &  0    &    $2.8 \pm 0.9$    &    $3 \pm 1$     &  $ 21 \pm 2$    &   $ 1.6 \pm 0.6$    &   $22 \pm 8$  \\
77  &      18:47:47.41   &   -1:54:25.60   &     $0.73 \times 0.45$   &    2000   &     8.9     &  0    &    $3.3 \pm 0.9$    &    $6 \pm 1$     &  $ 26 \pm 2$    &   $ 2.6 \pm 0.5$    &   $8 \pm 2$  \\
78  &      18:47:46.43   &   -1:54:29.81   &     $0.79 \times 0.45$   &    2200   &     7.7     &  0    &    $2.3 \pm 0.9$    &    $4 \pm 1$     &  $ 23 \pm 2$    &   $ 2.0 \pm 0.6$    &   $5 \pm 1$  \\
79  &      18:47:46.73   &   -1:54:25.63   &     $0.8 \times 0.46$   &    2300   &     11     &  0    &    $3 \pm 1$    &    $7 \pm 1$     &  $ 26 \pm 2$    &   $ 2.8 \pm 0.7$    &   $5 \pm 1$  \\
80  &      18:47:47.15   &   -1:54:24.66   &     $0.73 \times 0.45$   &    2000   &     7.2     &  0    &    $3 \pm 1$    &    $3 \pm 1$     &  $ 22 \pm 2$    &   $ 1.6 \pm 0.7$    &   $5 \pm 2$  \\
81  &      18:47:47.24   &   -1:54:18.47   &     $0.74 \times 0.46$   &    2100   &     8.4     &  0    &    $2.9 \pm 0.6$    &    $5.4 \pm 0.7$     &  $ 23 \pm 2$    &   $ 2.6 \pm 0.5$    &   $7 \pm 1$  \\
82  &      18:47:46.90   &   -1:54:33.98   &     $0.63 \times 0.45$   &    1600   &     7.7     &  0    &    $3 \pm 1$    &    $3 \pm 1$     &  $ 21 \pm 2$    &   $ 1.9 \pm 0.6$    &   $10 \pm 3$  \\
83  &      18:47:46.71   &   -1:54:24.19   &     $0.75 \times 0.45$   &    2100   &     7.9     &  0    &    $3 \pm 1$    &    $4 \pm 1$     &  $ 28 \pm 2$    &   $ 1.5 \pm 0.6$    &   $4 \pm 2$  \\
84  &      18:47:46.37   &   -1:54:21.28   &     $0.78 \times 0.45$   &    2200   &     9.7     &  0    &    $3.1 \pm 0.8$    &    $6 \pm 1$     &  $ 26 \pm 2$    &   $ 2.7 \pm 0.5$    &   $6 \pm 1$  \\
85  &      18:47:46.51   &   -1:54:26.42   &     $0.7 \times 0.5$   &    2200   &     8.3     &  0    &    $2.5 \pm 0.9$    &    $5 \pm 1$     &  $ 27 \pm 2$    &   $ 1.8 \pm 0.5$    &   $4 \pm 1$  \\
86  &      18:47:46.74   &   -1:54:20.81   &     $0.58 \times 0.45$   &    1400   &     7.4     &  0    &    $2.5 \pm 0.9$    &    $3.1 \pm 0.9$     &  $ 26 \pm 2$    &   $ 1.3 \pm 0.4$    &   $11 \pm 4$  \\
87  &      18:47:47.24   &   -1:54:16.06   &     $0.54 \times 0.45$   &    1200   &     7.5     &  0    &    $2.3 \pm 0.6$    &    $2.7 \pm 0.7$     &  $ 23 \pm 2$    &   $ 1.4 \pm 0.4$    &   $18 \pm 5$  \\
88  &      18:47:47.28   &   -1:54:38.27   &     $0.69 \times 0.45$   &    1900   &     7.4     &  0    &    $2.5 \pm 0.8$    &    $3.6 \pm 0.8$     &  $ 21 \pm 2$    &   $ 2.0 \pm 0.5$    &   $7 \pm 2$  \\
89  &      18:47:46.30   &   -1:54:28.76   &     $0.68 \times 0.45$   &    1800   &     7.3     &  0    &    $2.7 \pm 0.9$    &    $3.6 \pm 0.9$     &  $ 26 \pm 2$    &   $ 1.5 \pm 0.4$    &   $6 \pm 2$  \\
90  &      18:47:46.39   &   -1:54:18.14   &     $0.84 \times 0.45$   &    2400   &     7.2     &  0    &    $2.7 \pm 0.7$    &    $4.9 \pm 0.8$     &  $ 24 \pm 2$    &   $ 2.4 \pm 0.5$    &   $4.4 \pm 0.9$  \\
91  &      18:47:46.75   &   -1:54:33.30   &     $0.63 \times 0.45$   &    1600   &     7.1     &  0    &    $3 \pm 1$    &    $3 \pm 1$     &  $ 21 \pm 2$    &   $ 1.8 \pm 0.8$    &   $10 \pm 4$  \\
92  &      18:47:46.86   &   -1:54:12.13   &     $0.79 \times 0.47$   &    2300   &     7.9     &  0    &    $2.5 \pm 0.5$    &    $4.5 \pm 0.5$     &  $ 22 \pm 2$    &   $ 2.4 \pm 0.4$    &   $4.7 \pm 0.8$  \\
93  &      18:47:46.96   &   -1:54:40.86   &     $0.69 \times 0.45$   &    1900   &     7.0     &  0    &    $2.7 \pm 0.7$    &    $3.7 \pm 0.8$     &  $ 21 \pm 2$    &   $ 2.1 \pm 0.5$    &   $8 \pm 2$  \\
94  &      18:47:46.74   &   -1:54:37.80   &     $0.67 \times 0.45$   &    1800   &     7.0     &  0    &    $2.7 \pm 0.6$    &    $3.9 \pm 0.7$     &  $ 21 \pm 2$    &   $ 2.2 \pm 0.5$    &   $9 \pm 2$  \\
95  &      18:47:46.58   &   -1:54:25.67   &     $0.72 \times 0.45$   &    2000   &     7.1     &  0    &    $2.7 \pm 0.9$    &    $3.1 \pm 0.9$     &  $ 28 \pm 2$    &   $ 1.2 \pm 0.4$    &   $4 \pm 1$  \\
96  &      18:47:47.08   &   -1:54:23.54   &     $0.65 \times 0.45$   &    1700   &     7.4     &  0    &    $3 \pm 1$    &    $4 \pm 1$     &  $ 24 \pm 2$    &   $ 2.0 \pm 0.6$    &   $9 \pm 3$  \\
97  &      18:47:47.16   &   -1:54:10.15   &     $0.72 \times 0.45$   &    2000   &     7.6     &  0    &    $1.9 \pm 0.4$    &    $4.1 \pm 0.6$     &  $ 21 \pm 2$    &   $ 2.3 \pm 0.5$    &   $7 \pm 1$  \\
98  &      18:47:46.40   &   -1:54:28.37   &     $0.68 \times 0.52$   &    2200   &     7.0     &  0    &    $2.3 \pm 0.9$    &    $4 \pm 1$     &  $ 27 \pm 2$    &   $ 1.5 \pm 0.4$    &   $3 \pm 1$  \\
99  &      18:47:45.39   &   -1:54:39.13   &     $1.26 \times 0.55$   &    3900   &     8.5     &  0    &    $1.7 \pm 0.6$    &    $7.1 \pm 0.9$     &  $ 23 \pm 2$    &   $ 3.4 \pm 0.6$    &   $1.4 \pm 0.2$  \\
100  &      18:47:45.22   &   -1:54:40.46   &     $0.7 \times 0.47$   &    2100   &     6.8     &  0    &    $2.1 \pm 0.5$    &    $3.1 \pm 0.5$     &  $ 24 \pm 2$    &   $ 1.4 \pm 0.3$    &   $4.1 \pm 0.8$  \\
\hline
\end{tabular} 
}
   \end{table}

   \setcounter{table}{0}
   \begin{table}
 \begin{threeparttable}
       \small\footnotesize
   \caption*{}{\textbf{Supplementary Table 1} | Characteristics of the cores detected at 1.3\,mm in W43-MM1 (continued).}
   \makebox[\textwidth][c]{
\begin{tabular}{cccccccccccc}  
\hline    
\#    &       RA  &     Dec.   &      Size    &     FWHM        &      Sig     &  $C_{\rm line}$    &   $S^{\rm peak}_{\rm 1.3mm}$  &   $S^{\rm int}_{\rm 1.3mm}$    &   ${\bar T}_{\rm dust}$  &    $ M_{\rm core}$    &   ${\bar n}_{\rm H2}$\\
          &  [J2000]         &  [J2000]            &       [$'' \times ''$]      & [AU]    &       &   [$\%$]   &   [mJy/beam] &  [mJy]  &   [K]    &  [$\msun$] &  $ \times 10^7 \rm cm^{-3}$ \\
\hline   
101  &      18:47:47.30   &   -1:54:35.03   &     $0.59 \times 0.45$   &    1500   &     6.8     &  0    &    $2.8 \pm 0.7$    &    $4.1 \pm 0.8$     &  $ 21 \pm 2$    &   $ 2.3 \pm 0.6$    &   $17 \pm 4$  \\
102  &      18:47:47.15   &   -1:54:33.19   &     $0.7 \times 0.45$   &    1900   &     6.7     &  0    &    $2 \pm 1$    &    $2 \pm 1$     &  $ 21 \pm 2$    &   $ 1.2 \pm 0.6$    &   $4 \pm 2$  \\
103  &      18:47:46.74   &   -1:54:19.44   &     $0.51 \times 0.45$   &    1200   &     7.0     &  0    &    $2 \pm 1$    &    $2 \pm 1$     &  $ 26 \pm 2$    &   $ 1.1 \pm 0.5$    &   $15 \pm 6$  \\
104  &      18:47:46.80   &   -1:54:18.43   &     $0.61 \times 0.45$   &    1500   &     6.9     &  0    &    $2.2 \pm 0.8$    &    $2.9 \pm 0.9$     &  $ 24 \pm 2$    &   $ 1.3 \pm 0.4$    &   $9 \pm 3$  \\
105  &      18:47:46.50   &   -1:54:16.24   &     $0.46 \times 0.45$   &    1200   &     6.9     &  0    &    $1.9 \pm 0.6$    &    $1.9 \pm 0.6$     &  $ 23 \pm 2$    &   $ 1.0 \pm 0.3$    &   $13 \pm 4$  \\
106  &      18:47:46.40   &   -1:54:36.61   &     $0.63 \times 0.45$   &    1600   &     6.9     &  0    &    $2.3 \pm 0.6$    &    $3.6 \pm 0.7$     &  $ 21 \pm 2$    &   $ 2.0 \pm 0.5$    &   $11 \pm 3$  \\
107  &      18:47:47.24   &   -1:54:35.60   &     $0.57 \times 0.47$   &    1500   &     6.8     &  0    &    $2.6 \pm 0.7$    &    $3.9 \pm 0.7$     &  $ 21 \pm 2$    &   $ 2.2 \pm 0.5$    &   $15 \pm 3$  \\
108  &      18:47:45.32   &   -1:54:39.13   &     $0.69 \times 0.45$   &    1900   &     6.6     &  0    &    $1.9 \pm 0.9$    &    $2.6 \pm 0.9$     &  $ 24 \pm 2$    &   $ 1.2 \pm 0.5$    &   $4 \pm 2$  \\
109  &      18:47:46.60   &   -1:54:18.00   &     $0.67 \times 0.45$   &    1800   &     6.7     &  0    &    $2.3 \pm 0.9$    &    $4 \pm 1$     &  $ 25 \pm 2$    &   $ 1.7 \pm 0.5$    &   $7 \pm 2$  \\
110  &      18:47:47.16   &   -1:54:11.30   &     $0.65 \times 0.45$   &    1700   &     6.8     &  0    &    $2.1 \pm 0.6$    &    $3.4 \pm 0.6$     &  $ 21 \pm 2$    &   $ 1.9 \pm 0.4$    &   $9 \pm 2$  \\
111  &      18:47:46.80   &   -1:54:20.16   &     $0.7 \times 0.45$   &    1900   &     6.5     &  0    &    $2 \pm 1$    &    $3 \pm 1$     &  $ 25 \pm 2$    &   $ 1.4 \pm 0.6$    &   $5 \pm 2$  \\
112  &      18:47:46.30   &   -1:54:37.37   &     $0.66 \times 0.45$   &    1800   &     6.4     &  0    &    $2.3 \pm 0.5$    &    $2.7 \pm 0.5$     &  $ 22 \pm 2$    &   $ 1.5 \pm 0.3$    &   $7 \pm 2$  \\
113  &      18:47:47.03   &   -1:54:10.58   &     $0.64 \times 0.47$   &    1800   &     6.4     &  0    &    $1.5 \pm 0.5$    &    $2.2 \pm 0.6$     &  $ 21 \pm 2$    &   $ 1.2 \pm 0.4$    &   $5 \pm 2$  \\
114  &      18:47:46.91   &   -1:54:17.75   &     $0.53 \times 0.45$   &    1200   &     6.4     &  0    &    $1.9 \pm 0.5$    &    $2.2 \pm 0.5$     &  $ 22 \pm 2$    &   $ 1.1 \pm 0.3$    &   $16 \pm 4$  \\
115  &      18:47:46.76   &   -1:54:12.64   &     $0.77 \times 0.57$   &    2700   &     6.9     &  0    &    $2.4 \pm 0.2$    &    $5.3 \pm 0.2$     &  $ 22 \pm 2$    &   $ 2.8 \pm 0.4$    &   $3.5 \pm 0.5$  \\
116  &      18:47:46.18   &   -1:54:30.92   &     $0.62 \times 0.45$   &    1600   &     6.4     &  0    &    $2.2 \pm 0.6$    &    $3.5 \pm 0.7$     &  $ 23 \pm 2$    &   $ 1.7 \pm 0.4$    &   $10 \pm 2$  \\
117  &      18:47:46.67   &   -1:54:23.80   &     $0.97 \times 0.47$   &    2800   &     6.7     &  0    &    $3 \pm 1$    &    $6 \pm 1$     &  $ 29 \pm 2$    &   $ 2.3 \pm 0.6$    &   $2.6 \pm 0.6$  \\
118  &      18:47:47.04   &   -1:54:14.90   &     $0.72 \times 0.45$   &    2000   &     6.1     &  0    &    $1.6 \pm 0.7$    &    $2.5 \pm 0.7$     &  $ 23 \pm 2$    &   $ 1.2 \pm 0.4$    &   $4 \pm 1$  \\
119  &      18:47:47.37   &   -1:54:11.81   &     $0.54 \times 0.45$   &    1200   &     6.4     &  0    &    $1.8 \pm 0.5$    &    $2.4 \pm 0.6$     &  $ 21 \pm 2$    &   $ 1.4 \pm 0.4$    &   $18 \pm 5$  \\
120  &      18:47:46.80   &   -1:54:11.92   &     $0.76 \times 0.63$   &    2900   &     6.2     &  0    &    $2.5 \pm 0.4$    &    $5.7 \pm 0.4$     &  $ 22 \pm 2$    &   $ 3.0 \pm 0.5$    &   $2.9 \pm 0.5$  \\
121  &      18:47:47.00   &   -1:54:39.53   &     $0.74 \times 0.45$   &    2100   &     6.3     &  0    &    $1.7 \pm 0.8$    &    $2.4 \pm 0.9$     &  $ 21 \pm 2$    &   $ 1.3 \pm 0.5$    &   $4 \pm 1$  \\
122  &      18:47:44.76   &   -1:54:53.14   &     $0.59 \times 0.56$   &    2000   &     6.7     &  0    &    $1.3 \pm 0.4$    &    $3.1 \pm 0.6$     &  $ 23 \pm 2$    &   $ 1.5 \pm 0.3$    &   $4.4 \pm 0.9$  \\
123  &      18:47:46.99   &   -1:54:4.90   &     $0.85 \times 0.45$   &    2400   &     6.3     &  0    &    $1.7 \pm 0.4$    &    $3.2 \pm 0.5$     &  $ 21 \pm 2$    &   $ 1.8 \pm 0.4$    &   $3.3 \pm 0.7$  \\
124  &      18:47:47.37   &   -1:54:21.38   &     $0.84 \times 0.45$   &    2300   &     5.8     &  0    &    $2.4 \pm 0.7$    &    $5.0 \pm 0.8$     &  $ 25 \pm 2$    &   $ 2.2 \pm 0.4$    &   $4.1 \pm 0.8$  \\
125  &      18:47:47.47   &   -1:54:38.09   &     $0.63 \times 0.45$   &    1700   &     6.3     &  0    &    $2.2 \pm 0.5$    &    $3.3 \pm 0.6$     &  $ 23 \pm 2$    &   $ 1.6 \pm 0.4$    &   $9 \pm 2$  \\
126  &      18:47:44.57   &   -1:54:39.71   &     $0.83 \times 0.53$   &    2700   &     6.4     &  0    &    $1.5 \pm 0.5$    &    $4.3 \pm 0.8$     &  $ 23 \pm 2$    &   $ 2.1 \pm 0.5$    &   $2.6 \pm 0.6$  \\
127  &      18:47:46.25   &   -1:54:22.25   &     $1.02 \times 0.55$   &    3300   &     6.2     &  0    &    $1.9 \pm 0.6$    &    $5.8 \pm 0.9$     &  $ 27 \pm 2$    &   $ 2.4 \pm 0.4$    &   $1.6 \pm 0.3$  \\
128  &      18:47:47.29   &   -1:54:26.32   &     $0.88 \times 0.57$   &    3000   &     5.9     &  0    &    $1.7 \pm 0.9$    &    $4 \pm 1$     &  $ 23 \pm 2$    &   $ 2.0 \pm 0.6$    &   $1.7 \pm 0.5$  \\
129  &      18:47:46.56   &   -1:54:6.95   &     $0.87 \times 0.48$   &    2600   &     6.1     &  0    &    $1.5 \pm 0.4$    &    $3.7 \pm 0.5$     &  $ 23 \pm 2$    &   $ 1.8 \pm 0.3$    &   $2.6 \pm 0.5$  \\
130  &      18:47:46.37   &   -1:54:47.74   &     $0.98 \times 0.45$   &    2700   &     5.7     &  0    &    $1.9 \pm 0.4$    &    $4.5 \pm 0.5$     &  $ 21 \pm 2$    &   $ 2.5 \pm 0.5$    &   $3.0 \pm 0.5$  \\
131  &      18:47:46.55   &   -1:54:16.63   &     $1.11 \times 0.5$   &    3300   &     5.8     &  0    &    $1.6 \pm 0.6$    &    $5.1 \pm 0.8$     &  $ 23 \pm 2$    &   $ 2.6 \pm 0.5$    &   $1.7 \pm 0.3$  \\
\hline    
\end{tabular} 
}
\vskip 0.3cm
\begin{tablenotes}[para,flushleft]
\item Notes: RA, right ascension; Dec., declination; 
Size at half-maximum; FWHM, Size deconvolved with the $0.44''$ telescope beam; 
\item Sig, significance level of \emph{getsources} core extraction; 
line contamination ratio, $C_{\rm line}=[S^{\rm peak}{\rm (1.9\,GHz)}-S^{\rm peak}{\rm (65\,MHz)}]\;/\;S^{\rm peak}{\rm (65\,MHz)}$;
\item $S^{\rm peak}_{\rm 1.3mm}$ and $S^{\rm int}_{\rm 1.3mm}$, peak and total integrated flux corrected for line contamination and local background; 
${\bar T}_{\rm dust}$, mean dust temperature %
\item read in Fig.~3; 
$ M_{\rm core}$, core mass; 
mean density, ${\bar n}_{\rm H2}=  \frac{M_{\rm core}}{\frac{4}{3} \pi \, \mu \, m_{\rm H} \, ({\rm\it FWHM}/2)^{3}}$, where $\mu=2.8$ is the mean molecular weight per H$_2$ molecule 
\item  and $m_{\rm H}$ is the hydrogen mass.
\end{tablenotes}
\end{threeparttable}
   \end{table}

\setcounter{table}{1}
\vskip 0.5cm
   \begin{table}[htb]
 \begin{threeparttable}
    \caption*{}{\textbf{Supplementary Table 2} | Tests performed to evaluate the uncertainty of the reference CMF fit of Fig.~2b.}  
   \label{test}
   {\centering 
    \begin{tabular}{l | c c}
\hline 
 & Mass range & $\gamma$    \\
\hline
\textbf{Reference cumulative CMF of all cores extracted by \emph{getsources}}
& $>\!1.6\,\msun$ &  $ \bf -0.96 \pm 0.02$   \\
\textbf{\hskip 1cm with $5\sigma$ uncertainty derived from the mass uncertainties}	& $>\!1.6\,\msun$ &  $ \bf -0.96 \pm 0.13$   \\
\hskip 1cm low-mass regime & $1.6-20\,\msun$ & $-0.93\pm 0.02$  \\
\hskip 1cm (high-mass regime, 9 cores) & ($>\!20\,\msun$) & ($-1.3 \pm 0.3$) \\
\hskip 1cm with a lower completeness level & $>\!4.5\,\msun$ & $-0.99 \pm 0.04$ \\
CMF of the 94 most robust cores  & $>\!1.6\,\msun$  & $-0.90 \pm 0.02$ \\
\hline
CMF with core masses estimated in the optical thin approximation & $>\!1.6\,\msun$ & $-0.98 \pm 0.04$   \\
Differential CMF with all cores and default assumptions & $>\!1.6\,\msun$ &  $-0.90 \pm 0.06 $  \\
\hline
CMF built from cores extracted in a classic-cleaned image & $>\!1.6\,\msun$  &  $-1.10 \pm 0.05$    \\
\hskip 4.9cm  in a merged ($\rm 7\,m+12\,m$) image & $>\!1.6\,\msun$ or $>\!5\,\msun$  &  $-1.10 \pm 0.04$    \\
\hskip 4.9cm  with {\sc MRE-GaussClumps}
& $>\!1.6\,\msun$ or $>\!5\,\msun$  &  $-1.08 \pm 0.04 $    \\
\hline
\end{tabular}}
\vskip 0.3cm
\begin{tablenotes}[para,flushleft]
Notes: CMFs are fitted by power-laws of the form $N(>\!\log(M)) \propto M^{\gamma}$, except for the differential CMF where the power-law is $dN/d\log(M) \propto M^{\gamma}$. Several mass ranges are used to fit the CMFs of less-constrained core samples derived from the merged ($\rm 7\,m+12\,m$) image and the MRE-GaussClumps algorithm. Except when specified otherwise, all uncertainties given here are $1\sigma$.
\end{tablenotes}
 \end{threeparttable}
\end{table}

\end{document}